\documentclass[12pt]{article}

\renewcommand{\title}[1]{\begin{center}\bf\Large #1\end{center}}

\renewcommand{\author}[1]{\begin{center}\large #1\end{center}}

\usepackage{latexsym,amsfonts}
\newcommand{\rr}{\mathbb{R}}

\def\theequation{\arabic{section}.\arabic{equation}}

\begin{document}

\title{Poisson Structure and Moyal Quantisation of the Liouville
 Theory }
\author{
 George Jorjadze${}^a$
\footnote{email: \tt jorj@rmi.acnet.ge}
 and Gerhard Weigt${}^b$
\footnote{email: \tt weigt@ifh.de} \\
{\small${}^a$Razmadze Mathematical Institute,}\\
  {\small M.Aleksidze 1, 380093, Tbilisi, Georgia}\\
{\small${}^b$DESY Zeuthen, Platanenallee 6,}\\
{\small D-15738 Zeuthen, Germany}}

\begin{abstract}
\noindent
The symplectic and Poisson structures of the Liouville theory are
derived from the symplectic form of the $SL(2,\rr)$ WZNW theory by
gauge invariant Hamiltonian reduction. Causal non-equal time Poisson
brackets for a Liouville field are presented.  Using the symmetries of
the Liouville theory, symbols of chiral fields are constructed and
their $*$-products calculated. Quantum deformations consistent with
the canonical quantisation result, and a non-equal time commutator
is given.
\end{abstract}

\baselineskip=20pt

\vspace{0.3cm}

\setcounter{equation}{0}
\section{Introduction}

\noindent
Wess-Zumino-Novikov-Witten(WZNW) models \cite{WZNW} are fascinating
two-dimensional integrable conformal field theories which turn
up in many areas of physics and mathematics. A rich variety of additional
integrable theories is given by their cosets.  Hamiltonian
reduction \cite{Balog, FJW} provides a complete description of the
relationship between these theories.
In particular, the symplectic and Poisson
structures of this theory \cite{Goddard, Gawedzki} and its cosets can
be derived from the general solution of the equations of motions
of the WZNW model.
Moreover,  Hamiltonian reduction also proves
to be a general method for the integration of gauged WZNW
theories. It might be worthwhile to illustrate this by considering
the $SL(2,\rr)$ WZNW theory together with the Liouville theory
\cite{Thorn}-\cite{OW} and the $SL(2,\rr)/U(1)$ black hole model
\cite{BCR, FJW}.

\noindent
Our basic aim is to quantise the coset theories.
Even for these well-studied cases fundamental problems remain.
As an alternative to standard  canonical quantisation
\cite{Thorn, Neveu, OW, FJW} it seems
to be advantageous to consider a Moyal formalism
\cite{Weyl}-\cite{Berezin}. Such methods have recently been
variously applied to field theories with non-commutative geometry.
In our case, however, such
a formalism arises quite naturally as a consequence of a transitive
symmetry group acting on the phase space.
This symmetry may prove useful in the explicit construction
of Liouville correlation functions.

\noindent
In this paper we shall restrict our attention to Liouville theory.  As
well as being the simplest coset model, there is an extensive
literature on quantum Liouville theory \cite{Thorn}-\cite{OW} against
which we can test our Moyal approach.  Before we do the quantisation,
a full description of the classical Hamiltonian reduction will be
given.  The classical analogue of the exchange algebra is derived, and
from this we deduce the general causal non-equal time Poisson brackets
for a Liouville exponential. Such non-equal time Poisson structures
may play a role in other interacting field theories.

\noindent
In chapter $2$ we derive the basic Poisson brackets of the $SL(2,\rr)$
WZNW theory from its symplectic form and discuss the symmetry
properties of this theory. Chapter $3$ describes the nilpotent
reduction of these structures to the Liouville theory and presents in
particular the symmetries on the phase space and the causal non-equal
time Poisson brackets. After the introduction of a symbol calculus,
$*$-products of chiral fields and a non-equal time commutator are
calculated. Chapter $5$ summarises the results, and
some technical details are provided in four appendices.

\setcounter{equation}{0}
\section{The $SL(2,\rr)$ WZNW theory}

\noindent
The WZNW theory has a chiral structure and its investigation can
essentially be reduced to the analysis of the chiral or anti-chiral
part. The general solution of the dynamical equation gives the
$SL(2,\rr)$ field $g(\tau,\sigma)= g(z) \, \bar g(\bar z)$ as a product
of the chiral and anti-chiral fields $g(z)$ and $\bar g(\bar z)$,
 where $z=\tau +\sigma$, $\bar z=\tau -\sigma$ are
light cone coordinates. The symplectic form and the Hamiltonian have a
chiral structure as well.

\subsection{The basic Poisson brackets}

\noindent
We consider periodic boundary conditions in
$\sigma$ for the $SL(2,\rr)$ field $g(\tau,\sigma)$, and the chiral fields
have a monodromy $g(z+2\pi)=g(z)M$ with $M\in SL(2,\rr)$.  The
chiral part of the symplectic form of the $SL(2,\rr)$ WZNW theory
 is then \cite{Goddard, Gawedzki}
\begin{eqnarray}\label{omega-WZ}
\omega =-\frac{1}{\gamma^2}\,
\int_\tau^{\tau +2\pi}\,\, \langle(g^{-1}(z)\,\delta g(z))'\wedge\,
g^{-1}(z)\,\delta g(z)\rangle\,dz\\ \nonumber
-\frac{1}{\gamma^2}\,
\langle\delta M\, M^{-1}\, \wedge\, g^{-1}(\tau)\delta g(\tau)\rangle.
\end{eqnarray}
$\gamma >0$ is a coupling constant, $'$ denotes differentiation,
and the normalised trace is defined by $\langle
\cdot\rangle=-\frac{1}{2}\mbox{tr} (\cdot )$.

\noindent
Let us take the basis of the $sl(2,\rr)$ algebra
\begin{equation}\label{T}
  t_0=\left( \begin{array}{cr}
  0&-1\\1&0 \end{array}\right),~~~~
   t_1=\left( \begin{array}{cr}
  0&~1\\1&~0 \end{array}\right),~~~~
 t_2=\left( \begin{array}{cr}
  1&0\\0&-1 \end{array}\right).
\end{equation}
It satisfies the relations
$t_m~t_n=-\eta_{mn}~I+\epsilon_{mn}~^l~t_l$, where $I$ is the unit
matrix, $\eta_{mn}=\mbox{diag}(+,-,-)$ the metric tensor of a $3d$
Minkowski space, and $\epsilon_{012}=1$. For the matrices $t_n$ one has
$\langle t_m~t_n \rangle =\eta_{mn}$, $\langle t_l~ t_m~t_n \rangle =
\epsilon_{lmn}$
 and for any $a\in sl(2,\rr)$ \,\,
$t^n\,\langle t_n~a \rangle =a$. We shall also use the nilpotent
elements of the algebra $t_\pm =t_1\pm t_0$.

\noindent
In general, the 2-form (\ref{omega-WZ}) is singular and one has to
quotient the space of the chiral fields by $SL(2,\rr)$ transformations
$g(z)\mapsto g(z)N$ \cite{Goddard} under which $M\mapsto N^{-1}MN$. The
monodromy $M$ can be so transformed into an abelian subgroup, and we shall
choose
\begin{equation}\label{monodromy-g1}
  M=\left( \begin{array}{cr}
  e^{\lambda} & 0\\ 0 & e^{-\lambda} \end{array}\right),
~~~~\mbox{with}~~~~\lambda < 0,
\end{equation}
in order to get a regular Liouville field after Hamiltonian reduction.
The 2-form (\ref{omega-WZ}) then becomes symplectic and its inversion
gives the Poisson brackets (see Appendix A for the explicit
calculations)
\begin{eqnarray}\label{PB-g-lambda}
\{2\lambda ,\,g(z)\} = \gamma^2 \,g(z)\,t_2,\\
\label{PB-g}
\{\,g_{ab}(z),\, g_{cd}(y)\,\}=
\frac{\gamma^2}{4}\,[\,(\,g(z)\,t_2\,)_{ab}\,\,
(\,g(y)\,t_2\,)_{cd}\,\,\epsilon(z-y) \nonumber\\
+(\,g(z)\,t_-\,)_{ab}\,\,
(\,g(y)\,t_+\,)_{cd}
\theta_{-2\lambda} (z-y)\nonumber\\
+\, (\,g(z)\,t_+\,)_{ab}\,\,
(\,g(y)\,t_-\,)_{cd}
\theta_{2\lambda} (z-y)].
\end{eqnarray}
Here $\epsilon (z)$ is the stair-step function $\epsilon (z)= 2n+1~$
for $~2\pi n <\,z\,<\, 2\pi (n+1)$, and
\begin{equation}\label{theta_lambda}
\theta_{2\lambda} (z-y)=\frac{e^{\lambda \epsilon(z-y)}}{2\,\sinh\lambda}\,
\end{equation}
is the Green's function \cite{Thorn} which
inverts  the operator $\partial_z$ on
functions $A(z)$ with the monodromy $A(z+2\pi)=e^{2\lambda}A(z)$
for $\lambda\neq 0$.
Since in the interval $z\in(-2\pi, 2\pi)$ \, $\epsilon(z)=sign\,(z)$,
we have
\begin{equation}\label{theta_lambda1}
\theta_{2\lambda} (z)=\frac{\cosh\lambda}{2\sinh\lambda}+
\frac{1}{2}\,\epsilon (z).
\end{equation}
Equations (\ref{PB-g-lambda}), (\ref{PB-g}) are the basic Poisson
brackets of the $SL(2,\rr)$ WZNW theory.
The corresponding relations of the coset theories can be derived
through the Hamiltonian reduction.

\subsection{The Kac-Moody and conformal symmetries}

\noindent
As a consequence of the basic algebra (\ref{PB-g-lambda}), (\ref{PB-g})
the components of the (periodic) Kac-Moody currents
\begin{equation}\label{KM-current}
J_n (z)=\frac{1}{\gamma^2}\,\langle\,t_n\, g\,'(z)\, g^{-1}(z)\,\rangle
\end{equation}
commute with $\lambda$ and satisfy
\begin{equation}\label{PB-J-g}
\{J_n (z),\, g (y)\} = -\frac{1}{2}(t_n\, g (y))\,\delta (z-y),
\end{equation}
which means that the currents $J_n(z)$ are generators of left multiplications
\begin{equation}\label{KM-group}
R(f): \,g(z)\,\mapsto \, f(z)\,g(z)~~~~~(f(z)\in SL(2,\rr)).
\end{equation}
Since the symplectic form (\ref{omega-WZ}) contains the 1-forms
$g^{-1}(z)\,\delta g(z)$ which are invariant under left multiplications,
(\ref{KM-group}) are symmetry transformations of the system.
The $SL(2,\rr)$ Kac-Moody algebra follows directly from (\ref{KM-current})
and (\ref{PB-J-g}) as
\begin{equation}\label{KM-algebra}
\{J_m (z),\, J_n (y)\} = \epsilon_{mn}\,^l\,J_l (z)\,\delta (z-y)
+\frac{1}{2\gamma^2}\,\eta_{mn}\,\delta\,'(z-y),
\end{equation}
and for the chiral Sugawara energy momentum tensor
\begin{equation}\label{T-WZ}
T_g (z)= -\gamma^2\, J_n(z)\,J^n(z)
\end{equation}
we get the Poisson bracket relations
\begin{eqnarray}\label{PB-T-g}
&&\{T_g (z),\, g (y)\} =  g\,' (y)\,\delta (z-y),\\
\label{PB-T-J}
&&\{T_g (z),\, J_n (y)\} = J_n'(y)\,
\delta (z-y) -J_n(y)\delta\,' (z-y),\\
\label{PB-T-T}
&&\{T_g (z),\, T_g (y)\} = T_g\,'(y)\,\delta (z-y) -2T_g(y)\delta\,'(z-y).
\end{eqnarray}
We see that $g(z)$, $J_n(z)$ and
$T_g(z)$ have the conformal weights $0$, $1$ and $2$, respectively.
$T_g(z)$ generates the conformal transformations
$ g(z)\,\mapsto \,g(\xi(z))$,
which leave (\ref{omega-WZ}) invariant too. The symmetry of the
$SL(2,\rr)$ theory is therefore given by the semi-direct product of the
conformal and the Kac-Moody groups.

\noindent
In order to complete the description of the $SL(2,\rr)$ WZNW
structures by the anti-chiral part, we have to replace the left
invariant 1-form $g^{-1}(z)\,\delta g(z)$ of (\ref{omega-WZ}) by the
right invariant $\delta \bar g(\bar z)\,\bar g^{-1}(\bar z)$. So we
get the anti-chiral symplectic form of the theory with a similar
symmetry structure. The monodromy of the anti-chiral field
is given by $\bar g(\bar z-2\pi)=\bar M\bar g(\bar z)$ and the
periodicity of the WZNW field $g(\tau,\sigma)$ requires $\bar M= M
^{-1}$.

\setcounter{equation}{0}

\section {The Liouville theory}

The Liouville theory can be obtained by nilpotently gauging the
$SL(2,\rr)$ WZNW theory, imposing constraints for the chiral and
anti-chiral fields separately \cite{Balog}.   For the chiral part the
constraint is
\begin{equation}\label{constraint}
J_+(z)+\rho =0,~~~~~\mbox{where}~~~~
J_+(z)=\frac{1}{\gamma^2}\langle t_+\, g'(z)\, g^{-1}(z)\rangle.
\end{equation}
$\rho > 0$ is a fixed parameter and $t_+$ the nilpotent element
of the $sl(2,\rr)$ algebra. $J_+(z)+\rho=0$ is a first class constraint
which generates gauge transformations.

\noindent
In fact, we  apply a gauge invariant version
of the Hamiltonian reduction of \cite{FJW}
to the nilpotent case.

\subsection{The exchange algebra}

\noindent
Let us use for the variables and Poisson brackets of the
reduced system the same notation as for the $SL(2,\rr)$ WZNW theory.
Anticipating the Liouville theory we rename the gauge invariant
components of the chiral field $g_{11}(z)=\psi(z)$ and
$g_{12}(z)=\chi(z)$. Due to  gauge invariance
the form of the Poisson brackets
between $\psi$ and $\chi$ is obviously covariant under the
reduction.  Thus we can simply read off the result directly
from the basic Poisson brackets of the $SL(2,\rr)$ WZNW theory
(\ref{PB-g}) as
\begin{eqnarray}\label{PB-psi-psi}
&&\{\psi (z),\psi(y)\}=\frac{\gamma^2}{4}\psi (z)\,\psi(y)\epsilon(z-y),
\\
\label{PB-chi_1-chi_1}
&&\{\chi(z),\chi(y)\}=\frac{\gamma^2}{4}\chi(z)\,
\chi(y)\epsilon(z-y),\\
\label{PB-psi-chi}
&&\{\psi (z),\chi(y)\}=
-\frac{\gamma^2}{4}\psi (z)\,\chi(y)\epsilon(z-y)\\
 \nonumber
&&~~~~~~~~~~~~~~~~~~~
+{\gamma^2}\,\chi (z)\,\psi(y)\theta_{2\lambda}(z-y).
\end{eqnarray}
This is the basic Poisson algebra of the reduced
system. The corresponding quantum commutation relations are
nothing but the well-known exchange algebra \cite{Neveu}.
Below we shall show that eqs. (\ref{PB-psi-psi})-(\ref{PB-psi-chi})
together with the similar
anti-chiral relations, provide locality and canonicity
of the Liouville field.

\subsection{The nilpotent reduction of the $SL(2,\rr)$ theory}

\noindent
From the constrained Kac-Moody current
\begin{equation}\label{constrained-T-J}
J(z)= \left( \begin{array}{cr}
 -J_2(z)& \rho~~\\
-J_-(z)& J_2(z) \end{array}\right), ~~~~~~~~~
(J_-=J_1-J_0),
\end{equation}
and the relation
\begin{equation}\label{g-J}
g'(z)=\gamma^2J(z)g(z),
\end{equation}
we easily find the reduced chiral field
\begin{equation}\label{constrained-g}
  g(z)=\,\left( \begin{array}{cr}
 ~~\,\psi(z)~~~&\,\chi(z)~~~~~~~~\\
\beta^{-1}\psi\,'(z)+\rho^{-1} J_2(z)\psi(z) &
~\beta^{-1}\chi\,'(z)+\rho^{-1} J_2(z)\chi(z)
\end{array}\right),
\end{equation}
where $\beta =\rho\gamma^2$. The gauge freedom of $g(z)$ is
given  by
$J_2(z)$. Since $det~ g=1$,  $\psi(z)$ and $\chi(z)$ have a
constant Wronskian
\begin{equation}\label{wronskian}
\psi(z)\chi'(z)\,-\,\psi'(z)\chi(z)=\beta ,
\end{equation}
and (\ref{monodromy-g1}) defines  their
monodromy
\begin{equation}\label{monodromy-psi}
\psi(z+2\pi)=e^\lambda\,\,\psi(z),~~~~~
\chi(z+2\pi)=e^{-\lambda}\,\,\chi(z).
\end{equation}

\noindent
To get a regular Liouville field we assume $\psi (z)>0$. The relation
(\ref{wronskian}) can then be integrated by using (\ref{monodromy-psi}),
and $\chi(z)$ becomes
\begin{equation}\label{chi-psi}
\chi(z)=\,\beta\,\psi (z)\,\int_0^{2\pi}\,\,
dy\,\,\frac{\theta_{-2\lambda}\, (z-y)}{\psi^2(y)}\,\, ,
\end{equation}
which is positive since $\lambda <0$ (see
(\ref{monodromy-g1})). So the phase space of the reduced system can be
parameterised by the field $\psi(z)$ only.  Inserting the
field (\ref{constrained-g}) into the symplectic form (\ref{omega-WZ}),
the reduced
symplectic form can be written in terms of gauge invariant variables only
\begin{equation}\label{omega-psi}
\omega = \frac{1}{\gamma^2}\,
\int_\tau^{\tau +2\pi}\,\, \frac{1}{\psi^2(z)}
\,\delta\psi'(z)\wedge\,\delta\psi(z)\,dz +
\frac{1}{\gamma^2}\,
\delta\lambda \wedge\, \frac{\delta\psi(\tau)}{\psi(\tau)}.
\end{equation}
Since $\psi (z) >0$, we shall write $ \psi (z)=e^{-\gamma\phi(z)}$,
and (\ref{omega-psi}) gets in terms of $\phi(z)$ the free-field form
\begin{equation}\label{omega-phi}
\omega =
\int_\tau^{\tau +2\pi}\,\,
\,\delta\phi'(z)\wedge\,\delta\phi(z)\,dz +
\frac{\delta {p_0}}{2}\,
\wedge\, \delta\phi(\tau),
\end{equation}
with $p_0 =-{2\lambda}{/ \,\gamma}~$ (\,$p_0>0\,$). Eq. (\ref{monodromy-psi})
requires that this chiral field $\phi(z)$ has
the standard monodromy $\phi(z+2\pi)=\phi(z) +p_0/2$ of a free field.
Using the Fourier mode expansion
\begin{equation}\label{phi-mode}
\phi (z)= q_0+
\frac{p_0z}{4\pi}+\frac{i}{\sqrt {4\pi}}\sum_{n\neq 0}\frac{a_n}{n}\, e^{-inz},
\end{equation}
(\ref{omega-phi}) yields canonical Poisson brackets
for the modes
\begin{equation}\label{canonicalPB}
\{ p_0,\, q_0\} =1,~~~~~~~\{a_m,\, a_n\}=im\delta_{m+n,0}.
\end{equation}

\noindent
The classical vacuum configuration corresponds to $a_n=0$, whereas
the zero modes live on the half-plane $p_0>0$.
The vacuum is thus $(q,\, p)$-dependent and the corresponding fields
$\psi(z)$ and $\chi(z)$ are
\begin{equation}\label{vacuum}
\psi_0 (z,p)= e^{-\gamma\left (q+\frac{pz}{4\pi}\right)}\,,~~~~~~~~
\chi_0 (z,p)= \frac{2\pi\beta}{\gamma p}\,\,e^{\gamma\left (q
+\frac{ pz}{4\pi}\right)}\,.
\end{equation}

\noindent
In order to prove that the nilpotent gauging discussed here indeed leads
to the Liouville theory, we have to add the corresponding
anti-chiral part requiring the constraint $\bar J_+(\bar z)=\bar\rho$
with $\bar\rho >0$. $\bar J_+(\bar z)$ is defined similarly to
(\ref{constraint}) by the right Kac-Moody current. The gauge invariant
components are now $\bar\psi(\bar z)=\bar g_{12}(\bar z)$ and
$\bar\chi(\bar z)=\bar g_{22}(\bar z)$. They have the
monodromy
\begin{equation}\label{monodromy-psibar}
\bar\psi(\bar z-2\pi)=e^{-\bar\lambda}\,\,\bar\psi(\bar z),~~~~~
\bar\chi(\bar z-2\pi)=e^{\bar\lambda}\,\,\bar\chi(\bar z),
\end{equation}
and on the constrained surface they are related by the Wronskian
condition $\bar\psi(\bar z)\bar\chi'(\bar z)-
\bar\psi'(\bar z)\bar\chi(\bar z)=\bar\beta$, where $\bar\beta
=\gamma^2\bar\rho$. Assuming again $\bar\psi(\bar z)>0$ and `bosonise'
$ \bar\psi(\bar z)=e^{-\gamma\bar\phi(\bar z)}$
the reduced part is described correspondingly by
the anti-chiral free field $\bar\phi(\bar z)$ with the zero modes
$\bar p_0,\, \bar q_0,$  and the oscillators $\bar a_n$.

\noindent
The periodicity of the $SL(2,\rr)$ field $g(\tau, \sigma)$ requires
$\bar\lambda=\lambda$ which for the zero modes leads to the constraint
$p_0-\bar p_0=0$.  The free field $\phi(\tau,\sigma)=\phi(z)+
\bar\phi(\bar z)$ then becomes periodic as well and its canonical zero
modes are given by $q=q_0+\bar q_0$ and $p=p_0=\bar p_0 >0$. Note that this
constrained zero mode system which we got by reduction just corresponds to
the old-fashioned 'Fubini-Veneziano trick' to treat the dual model
zero modes conveniently. Indeed, we shall use this trick henceforth.
Then the chiral fields can be treated as independent fields with
closed Poisson brackets.

\noindent
For the periodic field ${g}(\tau ,\sigma)=g(z) \,\, \bar g(\bar
z)$ only the one component ${g}_{12}(\tau
,\sigma)=\psi(z)\bar\psi(\bar z)+\chi(z)\bar\chi(\bar z)$ remains
gauge invariant, and it is the Liouville field $\varphi(\tau, \sigma)$
which is expected to parameterise it
\begin{equation}\label{Liouville}
\psi(z)\bar\psi(\bar z)+\chi(z)\bar\chi(\bar z)=
e^{-\gamma \varphi(\tau, \sigma)}.
\end{equation}
From the constant Wronskian we obtain for
$\varphi(\tau, \sigma)$, indeed, the Liouville equation
\begin{equation}\label{Liouville-equation}
\left(\partial^2_\tau -\partial^2_\sigma\right)\varphi(\tau, \sigma)
+\frac{4m^2}{\gamma} e^{2\gamma\varphi(\tau, \sigma)}=0,~~~~\mbox{with}~~~~
m^2 =\beta\bar\beta,
\end{equation}
and eqs. (\ref{Liouville}) and (\ref{chi-psi}) provide its
general solution in the standard parametrisation
\begin{equation}\label{Liouville-sol}
e^{2\gamma\varphi(\tau, \sigma})=
\frac{A'(z)\bar A'(\bar z)}{[1+m^2\,A(z)\bar A(\bar z)]^2},
\end{equation}
where
\begin{equation}\label{A(z)}
A(z)=\int_0^{2\pi}\,\,
dy\,\,\theta_{\gamma p}\, (z-y)\,e^{2\gamma\phi(y)}.
\end{equation}
$\varphi(\tau,\sigma)$ is periodic, and one can show that this
description covers the class of all regular periodic Liouville fields.

\noindent
It is worth mentioning that the integration of
both the Liouville and the $SL(2,\rr)/U(1)$ theory \cite{FJW} by
Hamiltonian reduction is a strong indication that this approach can be
generalised to any gauged WZNW theory.
The standard Lax-pair method \cite{LS}
applies to the nilpotent gaugings only.

\subsection{The symmetries of the Liouville theory}

\noindent
The symmetry properties of the chiral fields $\psi(z)$ and
$\chi(z)$ will play an important role to implement the Moyal
deformation quantisation.
Putting  (\ref{constrained-T-J}) and
(\ref{constrained-g}) into
(\ref{g-J}), comparison of the components  yields
\begin{equation}\label{schrodinger-eq}
\frac{\psi''(z)}{\psi(z)}=\frac{\chi''(z)}{\chi(z)}=
\gamma^2 (T_g(z)-J_2'(z)).
\end{equation}
$T_g(z)=\gamma^2(J_2^2(z)-\rho J_-(z))$ is the reduced
energy-momentum tensor of (\ref{T-WZ}) which becomes gauge invariant
only through the `improvement' $T_g(z) \mapsto T(z) =T_g(z) -J'_2(z)$
\cite{JPP},  we have  got in eq. (\ref{schrodinger-eq}).
From (\ref{PB-T-g})-(\ref{PB-T-T}) we easily find
\begin{eqnarray}\label{PB-T-psi}
&&\{T (z),\psi(y)\}=\psi\,'(y)\,\delta (z-y) +\frac{1}{2}
\psi(y)\delta\,'(z-y),\\
\label{PB-T-psi1}
&&\{T (z),\chi(y)\}=\chi\,'(y)\,\delta (z-y) +\frac{1}{2}
\chi(y)\delta\,'(z-y),\\
\label{PB-T-T,L}
&&\{T (z), T(y)\}=
T\,'(y)\,\delta (z-y) -2T(y)\delta\,'(z-y)\\
\nonumber
&&~~~~~~~~~~~~~~~~~~~+\frac{1}{2\gamma^2}\delta\,'''(z-y).
\end{eqnarray}
We see that both $\psi(z)$ and $\chi(z)$ have the same
conformal weight $-\frac{1}{2}$ and that the algebra of the improved
tensor is deformed by a central extension only.  Thus we have, as
expected, conformal invariance of the Liouville theory, and the gauge
invariant $T(z)$ will therefore be identified with its energy-momentum
tensor.

\noindent
The relation (\ref{schrodinger-eq}) corresponds to the Schr\"odinger
equation, well-known in Liouville theory, for both $\psi(z)$ and
$\chi(z)$ with the solutions $\psi(z)=e^{-\gamma\phi(z)}$ and
(\ref{chi-psi}).  It allows us to transform the energy-momentum tensor
into a free-field form with the typical improvement term
\begin{equation}\label{T-phi}
T(z) =\phi'\,^2\,(z)-\frac{1}{\gamma}\,\phi''\,(z).
\end{equation}

\noindent
Our deformation quantisation will rely heavily on the following
important observation.  The symplectic form (\ref{omega-psi}) is
invariant under $\psi(z)\mapsto e^{-\gamma\rho(z)}\psi(z)$ with
periodic $\rho(z)$.  These transformations are generated by
$\phi\,'(z)$
\begin{eqnarray}\label{PB-phi'-psi}
\{\phi'(z),\, \psi(y)\}=-\frac{\gamma}{2}\delta(z-y)\psi(y),
\end{eqnarray}
and for the chiral free field this corresponds to the
translations $\phi(z)\mapsto \phi(z)+\rho(z)$. These
translations and the conformal transformations represent together
the symmetry
group of the free-field symplectic form (\ref{omega-phi}). The
corresponding Lie algebra is given by (\ref{PB-T-T,L}) and
\begin{eqnarray}\label{PB-phi'-phi'}
&&\{\phi'(z),\, \phi'(y)\}=-\frac{1}{2}\delta\,'(z-y),\\
\label{PB-T-phi'}
&&\{ T (z),\, \phi' (y)\}= \phi\,'' (y)\delta(z-y)-
\phi\,' (y)\delta\,'(z-y)+\frac{1}{2\gamma}\delta\,''(z-y).
\end{eqnarray}
Comparing this with (\ref{KM-group}), (\ref{KM-algebra}) and
(\ref{PB-T-J}), we see that the $\phi'(z)$ and the Kac-Moody current
obviously play a similar role for the Liouville and $SL(2,\rr)$ WZNW
theories, respectively.

\noindent
Although the fields $\psi(z)$ and $\chi(z)$ have the same conformal
weight their transformations with respect to the translation group is
different. In distinction to
(\ref{PB-phi'-psi}) the infinitesimal translation of $\chi(z)$
\begin{eqnarray}\label{PB-phi'-chi}
\{\phi'(z),\, \chi(y)\}=-\frac{\gamma}{2}\delta(z-y)\chi(y)+
\gamma\beta\, \theta_{\gamma p}(y-z)\,e^{-\gamma\phi(y)}e^{2\gamma\phi(z)}
\end{eqnarray}
produces the  bilocal field
\begin{equation}\label{bilocal}
B(y,z)=\theta_{\gamma p}(y-z)\,e^{-\gamma\phi(y)}e^{2\gamma\phi(z)}\,.
\end{equation}
This bilocal field transforms linearly with respect to the translation
\begin{equation}\label{bilocal1}
B(y,x)\mapsto e^{-\gamma\rho(y)}\,e^{2\gamma\rho(x)}\,B(y,x),
\end{equation}
and its $x$-integration gives
\begin{equation}\label{B-chi}
\chi (y)=\beta\int_0^{2\pi}\,dx\, B(y,x).
\end{equation}

\noindent
Since $p>0$, there is no global translation symmetry in the $p$
direction of the half-plane. In order to have  a group
with a transitive action on the phase space at hand, we introduce the
dilatations $p\mapsto e^{-\varepsilon} p$, $q\mapsto e^{\varepsilon}q$
generated by $K=pq$. The vacuum configuration (\ref{vacuum})
transforms for $q=0$ under these dilatations as
\begin{equation}\label{vacuum-symm}
\psi_0(z,e^{-\varepsilon}\,p)=\psi_0(e^{-\varepsilon}\,z,p),~~~~~~
\chi_0(z,e^{-\varepsilon}\,p)=e^{\varepsilon}\,\chi_0(e^{-\varepsilon}\,z,p).
\end{equation}
Adding this one-parameter group to the translations $\phi(z)\mapsto
\phi(z)+\rho(z)$ the new group $G(\rho(z),\varepsilon)$ is defined.
The Moyal quantisation of the Liouville
theory is based just on this symmetry group.

\subsection{The non-equal time Poisson structure of Liouville fields}

\noindent
In this section we calculate non-equal time Poisson brackets for the
exponential of the Liouville field (\ref{Liouville}) denoted here by
\begin{equation}\label{u}
u(z,\bar z)=
e^{-\gamma \varphi(\tau, \sigma)}.
\end{equation}
Using the exchange algebra
(\ref{PB-psi-psi})-(\ref{PB-psi-chi}) together with the similar
 anti-chiral relations,  we easily find
\begin{eqnarray}\label{PB-Liouville1}
\{u(z, \bar z),\, u(y, \bar y)\}=\gamma^2\,\Theta_{\gamma p}\,
\psi(z)\chi(y)\bar\chi(\bar z)\bar\psi(\bar y)+
\gamma^2\,\Theta_{-\gamma p}\,\chi(z)\psi(y)\bar\psi(\bar z)\bar\chi(\bar y)
\nonumber\\
+\frac{\gamma^2}{2}\,\Theta\,
[\psi(z)\psi(y)\bar\psi(\bar z)\bar\psi(\bar y)
+\chi(z)\chi(y)\bar\chi(\bar z)\bar\chi(\bar y)
\nonumber\\
-\psi(z)\chi(y)\bar\psi(\bar z)\bar\chi(\bar y)
-\chi(z)\psi(y)\bar\chi(\bar z)\bar\psi(\bar y)].
\end{eqnarray}
Here we have introduced the notation
\begin{equation}\label{Theta}
\Theta_{\gamma p}=\theta_{\gamma p}(z-y)+\theta_{-\gamma p}(\bar z-\bar y),
~~~~~~\Theta = \frac{1}{2}[\epsilon(z-y)+\epsilon(\bar z-\bar y)].
\end{equation}
With the relation
\begin{eqnarray}\label{uu-uu}
u(z, \bar z)\, u(y, \bar y)- u(z, \bar y)\, u(y, \bar z)
=\psi(z)\chi(y)\bar\psi(\bar z)\bar\chi(\bar y)
+\chi(z)\psi(y)\bar\chi(\bar z)\bar\psi(\bar y) \nonumber \\
-\psi(z)\chi(y)\bar\chi(\bar z)\bar\psi(\bar y)-
\chi(z)\psi(y)\bar\psi(\bar z)\bar\chi(\bar y),
\end{eqnarray}
which follows from (\ref{Liouville}) and the definition (\ref{u}), the
Poisson bracket (\ref{PB-Liouville1}) becomes
\begin{eqnarray}\label{PB-Liouville2}
\{u(z, \bar z),\, u(y, \bar y)\}=\gamma^2\,(\Theta_{\gamma p}-\Theta)
\psi(z)\chi(y)\bar\chi(\bar z)\bar\psi(\bar y)\nonumber\\
+\gamma^2\,(\Theta_{-\gamma p}-\Theta)
\,\chi(z)\psi(y)\bar\psi(\bar z)\bar\chi(\bar y)\nonumber\\
+\frac{\gamma^2}{2}
\Theta[2u(z, \bar y)\, u(y, \bar z) -u(z, \bar z)\, u(y, \bar y)].
\end{eqnarray}
In the fundamental interval $ z-y\in (-2\pi,\,2\pi)$ and $\bar z-\bar
y\in (-2\pi,\,2\pi)$, where (\ref{theta_lambda1}) gives
$\Theta_{\gamma p}=\Theta=\Theta_{-\gamma p}$, (\ref{PB-Liouville2})
reduces to the following non-equal time bracket relation
\begin{eqnarray}\label{PB-Liouville3}
\{u(z, \bar z),\, u(y, \bar y)\}=
\frac{\gamma^2}{4}[\epsilon(z-y)+\epsilon(\bar z-\bar y)]
\,[2u(z, \bar y)\, u(y, \bar z) -u(z, \bar z)\, u(y, \bar y)].
\end{eqnarray}
To our knowledge this result is not in the existing
literature. The step character of the function $\epsilon (z)$
provides causality of this Poisson bracket, and in particular its
equal time form vanishes.  For the Liouville theory on
the line where we do not have zero modes eq. (\ref{PB-Liouville3})
is valid in general.  It shows that the Poisson
brackets of the Liouville field at two different space-time points
$(z,\bar z),\,\,(y,\bar y)$ is expressed by the field in
these two and two other points $(z,\bar y),\,\,(y,\bar z)$ obtained by
exchanging the light-cone coordinates.

\noindent
When \,$ z-y$\, and \,$\bar z-\bar y$ \, are out of the
fundamental domain we still can get a closed form of
(\ref{PB-Liouville2}) in terms of the field $u$ .  For this purpose we
introduce two new space time points $(z +2\pi, \bar y +2\pi)$ and $(y
+2\pi,\, \bar z +2\pi)$ which are shifted by $2\pi$ (in time) with
respect to $(z,\,\bar y )$ and $(y,\, \bar z)$, respectively.  The
monodromies of $\psi$ and $\chi$ give
\begin{eqnarray}\label{Mon-u1}
u(z +2\pi, \bar y +2\pi)=e^{-\gamma p}\psi(z)\psi(\bar y)
+e^{\gamma p}\chi(z)\chi(\bar y),
\end{eqnarray}
which lead to
\begin{eqnarray}\label{z-bary}
\psi(z)\bar \psi(\bar y)=\frac{e^{\gamma p}\,u(z, \bar y)-
u(z +2\pi, \bar y +2\pi)}
{2\sinh \gamma p},\nonumber \\
\chi(z)\bar \chi(\bar y)=\frac{u(z +2\pi, \bar y +2\pi)
-e^{-\gamma p}\,u(z, \bar y)}
{2\sinh \gamma p}.
\end{eqnarray}
Similarly the monodromy of $u(y, \bar z)$ defines the products
$\psi(y)\bar \psi(\bar z)$ and $\chi(y)\bar \chi(\bar z)$. Inserting
these formulas into (\ref{PB-Liouville2}) we obtain the
non-equal time Poisson bracket for
$u(z, \bar z)$ in general. It now relates quadratic terms of $u$ at six
different space-time points.

\noindent
In the space-time coordinates
\begin{equation}\label{t,sigma}
\tau =\frac{1}{2}(z+\bar z),~~~~\sigma =\frac{1}{2}(z-\bar z),~~~~
\tau_1 =\frac{1}{2}(y+\bar y),~~~~\sigma_1 =\frac{1}{2}(y-\bar y),
\end{equation}
the fundamental interval corresponds to the time-interval
$|\tau-\tau_1|<2\pi$, where (\ref{PB-Liouville3}) is valid.
Outside of it we have $|\tau-\tau_1|>2\pi$, and the additional
terms of (\ref{PB-Liouville2}) contribute due to the
spatial periodicity.

\noindent
Differentiating (\ref{PB-Liouville3}) with respect to
$\tau $ and putting $\tau_1=\tau$ we obtain the equal time relation
\begin{equation}\label{canonical-PB}
\{ \partial_\tau e^{-\varphi(\tau, \sigma)},\,e^{-\varphi (\tau, \sigma_1)}\}=
\delta(\sigma-\sigma_1)\, e^{-2\varphi(\tau, \sigma)},
\end{equation}
which is equivalent to the canonical Poisson bracket \,
$\{ \dot\varphi(\tau, \sigma),\,\varphi (\tau, \sigma_1)\}=
\delta(\sigma-\sigma_1)$.

\noindent
With these results we could also calculate the Poisson bracket for
the physically more interesting arbitrary exponential of $\varphi
(\tau, \sigma_1)$, the Liouville vertex function.

\setcounter{equation}{0}
\section{A Moyal quantisation of the Liouville theory}

Following the ideas of deformation quantisation (for a general
treatment see \cite{Flato, Reuter}), we describe the quantum Liouville
theory not in the operator formalism but rather by functionals on the
phase space. In our case the formalism is based on the symmetry group
$G(\rho(z),\varepsilon )$ of Chapter $3.3$. Let us introduce the
elements we shall need  for our alternative quantisation of
the Liouville theory, discussing first a single oscillator.

\subsection{ The Moyal formalism for the oscillator and zero modes}

We consider the Hamiltonian
\begin{equation}\label{Hamilton1}
H=\frac{1}{2}\left(P^2 + \nu^2 Q^2 \right),
\end{equation}
and recall some well-known facts of quantum mechanics
\cite{Perelomov}.  The coherent states $|{\bf a}\rangle =
|P,Q; \nu\rangle$ with given frequency $\nu>0$ are labelled by the points
of the phase space.  The state $|{\bf a}\rangle$ is related to the
vacuum state $|\bf {0}\rangle$ by the Weyl group transformation
\begin{equation}\label{Weilgroup}
|{\bf a}\rangle =
\exp\frac{i}{\hbar}\left(\hat Q\, P-\hat P\,Q\right)
 |{\bf {0}}\rangle =
\exp\frac{1}{\hbar\nu}\left(\hat a^+\, a-\hat a\,a^*\right)
 |{\bf {0}}\rangle,
\end{equation}
which means that the translations of the coherent states
 are given up to a phase factor by the classical law.
$\hat a^+$, $\hat a$ are creation and annihilation
operators  with the canonical commutation relation \,
$[\hat a, \hat a^+]=\nu \hbar$, and
\begin{equation}\label{a,a+}
a=\frac{P-i\nu Q}{\sqrt{2}},~~~~~~
a^*=\frac{P+i\nu Q}{\sqrt{2}}
\end{equation}
are the corresponding classical variables.
The coherent states are eigenstates of the annihilation operator
\begin{equation}\label{eigenstate}
\hat a\,|{\bf a}\rangle = a\,|{\bf a}\rangle,
\end{equation}
and they are complete
\begin{equation}\label{I}
\int\, |{\bf a}\rangle\, d^2{\bf a}\, \langle{\bf a}|=\hat I,~~~~
\mbox{with}~~~d^2{\bf a}=(2\pi\hbar)^{-1}\,dp\,dq.
\end{equation}
As a consequence of the Baker-Hausdorf formula the scalar
product of the coherent states becomes
\begin{equation}\label{ba}
\langle {\bf b}\,|\,{\bf a}\rangle =\exp\left (-\frac{|a|^2+|b|^2 -2b^*\,a}
{2\hbar\nu} \right).
\end{equation}
The time evolution of the coherent states also follows the classical law
\begin{equation}\label{cl-law}
e^{-\frac{i}{\hbar}\hat H\,t}\,| a,\,\,a^*\rangle=
e^{-\frac{i}{2}\nu t}\,|e^{-i\nu t} a,
\,\,e^{i\nu t}a^* \rangle,
\end{equation}
It is an important observation that there is a one-to-one correspondence
between an operator  $\hat A$ on the Hilbert space and a
function $\check A(P,Q)$ on the phase space.
We specify such a map by the normal ordering prescription
\begin{equation}\label{symb}
\check A(a^*,a)=\langle {\bf a} |\hat A|{\bf a}\rangle,~~~~~~
\hat A =:\check A(\hat a^*,\hat a):
\end{equation}
The function $\check A(a^*,a)$ is known as the normal (Wick)
or Berezin symbol of the operator $\hat A$ \cite{Berezin}.
The non-diagonal matrix elements are given by
\begin{equation}\label{bAa}
\langle {\bf b}\,|\hat A|\,{\bf a}\rangle = \check A(b^*,a)
\,\langle {\bf b}\,|\,{\bf a}\rangle .
\end{equation}

\noindent
The non-commutativity of quantum mechanics is introduced through
the Moyal $*$-product of symbols
\cite{Moyal}, which is a symbol of the product of the
corresponding two operators
\begin{equation}\label{Moyalproduct}
\check A * \check B= \langle{\bf a}|\hat
A\,\hat B|{\bf a}\rangle.
\end{equation}
Using the completeness relation (\ref{I})
as well as (\ref{ba}) and  (\ref{bAa})
we find the useful integral representation
\begin{eqnarray}\label{*pr}
\check A * \check B =
\int\,d\mu(\xi) \,\,e^{-\frac{|\xi|^2}{\nu}}\,\, \check A(a^*,\,a+\sqrt{\hbar}\,\,\xi)
\, \check B(a^*+\sqrt{\hbar}\,\xi^*,\, a),
\end{eqnarray}
where we have  first made a shift of the integration
variables (\ref{I}) and then a dilatation
by $\sqrt\hbar~$. $~d\mu(\xi)=(2\pi)^{-1} dx dy $
is the corresponding normalised measure and $\xi=(x+i\nu
y)/{\sqrt{2}}$. Expanding the integrand in powers of $\hbar$, after
Gaussian integration, the $*$-product is easily seen to be a
deformation of the product of two ordinary functions
\begin{eqnarray}\label{*pr_hbar}
\check A* \check B= \check A\cdot \check B+
\hbar\nu \frac{\partial \check A}{\partial a}\frac{\partial \check B}
{\partial a^*}+
\frac{1}{2!}(\hbar\nu)^2 \frac{\partial^2 \check A}
{\partial a^2}\frac{\partial^2 \check B}{\partial a^{*\,2}}+\cdot\cdot\cdot .
\end{eqnarray}
The key object which corresponds to the commutator is the Moyal
$*$-bracket of two symbols
\begin{eqnarray}\label{*br}
\{\check A,\,\check B\}_*=\frac{i}{\hbar}(\check A*\check B-
\check B*\check A).
\end{eqnarray}
Since the $*$-product (\ref{*pr_hbar}) is associative, the Moyal
bracket obeys the Jacobi identity.
Equation (\ref{*pr_hbar}) gives the practically useful expansion
\begin{equation}\label{*brh}
\{\check A,\,\check B\}_*= \{\check A,\,\check B\} +
i\hbar\frac{\nu^2}{2!}\left(  \frac{\partial^2 \check A}{\partial a^2}
\frac{\partial^2 \check B}{\partial a^{*\,2}}-
\frac{\partial^2 \check B}{\partial a^2}
\frac{\partial^2 \check A}{\partial a^{*\,2}}
\right)+\cdot\cdot\cdot,
\end{equation}
where $\{\check A,\,\check B\}$ is the standard Poisson bracket.

\noindent
Applying these quantisation rules to the generators
of the Weyl transformations $\hat P$, $\hat Q$, and to the Hamiltonian $\hat
H=\hat a^+ \hat a$ the belonging symbols obviously coincide with their
classical counterparts, and their $*$-brackets with an
arbitrary function $\check A(P,Q)$ are simply the Poisson brackets. Such
non-deformation properties remain valid in general
for the generators of symmetry groups which define the coherent states
\cite{Perelomov}.

\noindent
For the Liouville theory the chiral part of the Hamiltonian
\begin{equation}\label{H_L}
H =\int_0^{2\pi}dz\,\tilde T(z)=\frac{p^2}{8\pi}+\sum_{n>0}|a_n|^2
\end{equation}
has for the non-zero modes the discussed oscillator structure. These
modes have therefore the standard coherent states with the frequency
parameter $\nu_n=n$ and the symbol calculus can be applied as before.

\noindent
But the treatment of the zero modes is essentially different since they
are given on the half-plane $p>0$ only. Here the Weyl
symmetry fails and the corresponding coherent states and symbols have
to be constructed
separately (see Appendix B).
On the half-plane the coordinate representation
does not exist, but we can still
make use of the
momentum representation on the Hilbert space $L^2(\rr_+)$.
The operator $\hat p$  then acts as a multiplication on a wave function
$\Psi (p)\in L^2(\rr_+)$, and the generator of dilatations  is
$\hat K=i\hbar p\partial_p +i\hbar/2$.
The operator $\hat q=i\hbar\partial_p$ is not self-adjoint on
$L^2(\rr_+)$,  its exponent  $e^{\beta\hat q}$ acts by
$e^{\beta\hat q}\,\Psi (p)=\Psi (p+i\hbar \beta)$
which leads to $A(\hat p)\,e^{\alpha\hat q}=
e^{\alpha\hat q}\,A(\hat p-i\hbar \alpha)$.
We shall associate the function $e^{2\alpha q}A(p)$ with the symbol of the
operator
\begin{equation}\label{hatA}
\hat A=e^{\alpha\hat q}A(\hat p)e^{\alpha\hat q}.
\end{equation}
Such symbols are just characteristic for our chiral fields.
The $*$-product
of two symbols then becomes
\begin{eqnarray}\label{*pr0}
{\check A} *{\check B} = e^{2(\alpha+\beta)q}\,\,
{ A}(p-i\hbar \beta)
\,\,{B}(p+i\hbar \alpha).
\end{eqnarray}
Together with (\ref{*pr}), this defines the $*$-product of symbols for
the Liouville theory.  From (\ref{*pr0}), it is easy to see that the
$*$-brackets of $p$ and $p^2$ with any $\check A$ are the Poisson
brackets.  The same is expected for the dilatation $K=pq$. Although
$K$ does not belong to the class $e^{2\beta q}B(p)$ its $*$-bracket
can be inferred by differentiating the $*$-bracket $\{e^{2\beta q}p, \check
A\}_*$ with respect to $\beta$ and putting $\beta =0$. Indeed, this
leads to $\{{\check A},\,K\}_*= \{{\check A},\,K\}$. In this manner
the $*$-brackets of other polynomials can be derived as well.

\subsection{The Moyal formalism for the Liouville theory}

For the Liouville theory the chiral symbols are functionals of the chiral
field $\phi(z)$, ${\check A}={\check A}[\phi(z)]$. In order to
specify them we make use of the
decomposition (\ref{phi-mode})
\begin{equation}\label{phi-decomp}
\phi (z)= q+
\frac{pz}{4\pi}+\phi^{+}(z)+\phi^{-}(z),~~~~\mbox{where}~~~~
\phi^{\pm}(z)= \pm \frac{i}{\sqrt{4\pi}}\sum_{n>0}
\frac{a_{\pm n}}{n}e^{\mp inz}.
\end{equation}
We shall treat here symbols containing the $q$ zero mode as
exponentials only
\begin{equation}\label{symbol}
{\check A[\phi(z)]}=e^{2\alpha q}\,{ A_0}[p,\,\phi^-(z),\phi^+(z)],
\end{equation}
and associate it with the operator
\begin{equation}\label{op-symbol}
{\hat A}=
e^{\alpha\hat q}\,:{A_0}[p,\,\hat\phi^-(z),\hat\phi^+(z)]:e^{\alpha\hat q}.
\end{equation}
Here $\,\,:~\,\,:\,\,$ denotes normal ordering of the oscillator
modes of $ A_0$.
Then the $*$-product of two symbols becomes
\begin{eqnarray}\label{*-pr}
{\check A} *{\check B} = e^{2(\alpha+\beta )q}\,\,
\int\,\prod_{n>0} d\mu(\xi_n) \,\,e^{-\frac{|\xi_n|^2}{n}}\times\,\,
~~~~~~~~~~~~~~~~~~~~~~~~~~~~~~~~~~~~~~~~~~~~~~
\nonumber \\
{ A_0}[p-i\hbar \beta,\,\phi^-(z),\,\phi^+(z)+\sqrt{\hbar}\,\,\xi^+(z)]
\,\,{ B_0}[p+i\hbar \alpha,\,\phi^-(z)+\sqrt{\hbar}\,\,\xi^-(z),\, \phi^+(z)],
\end{eqnarray}
where $d\mu(\xi_n)=(2\pi)^{-1}\,dp_n\,dq_n$ is the Liouville
measure on a plane, $\xi_n$ are the complex coordinates on it
\begin{equation}\label{xi-pm}
\xi_n=\frac{p_n-inq_n}{\sqrt 2},
~~~~\mbox{and}~~~~\xi^{\pm}(z)= \pm \frac{i}{\sqrt{4\pi}}\sum_{n>0}
\frac{\xi_{\pm n}}{n}\,e^{\mp inz}.
\end{equation}
Expanding the $*$-product (\ref{*-pr}) in powers of $\hbar$ we find
\begin{eqnarray}\label{*prh}
{\check A}*{\check B}={\check  A}\cdot{\check  B}
-\frac{i\hbar}{2}\left(\frac{\partial\check  A}{\partial p}
\frac{\partial\check B}{\partial q}-
\frac{\partial\check A}{\partial q}\ \, \frac{\partial \check B}{\partial p}+2i\,
\sum_{n>0}n\,  \frac{\partial \check A}
{\partial a_n}\,\,\frac{\partial \check B}
{\partial a^{*}_n}\right)\nonumber \\
-\frac{\hbar^2}{8}\left(\frac{\partial^2 \check A}
{\partial p^2}\,\frac{\partial^2 \check B}{\partial q^2}
+\frac{\partial^2 \check A}{\partial q^2}\,
\frac{\partial^2 \check B}{\partial p^2}
-2\frac{\partial^2 \check A}{\partial p \partial q}\,\frac{\partial^2 \check B}
{\partial p \partial q}
-4\sum_{n,m>0}mn\,\frac{\partial^2 \check A}
{\partial a_m\,\partial a_n}\,\,\frac{\partial^2 \check B}
{\partial a^{*}_m\,\partial a^{*}_n}\right)
\nonumber\\
-\frac{i\hbar^2}{2}\sum_{n>0}n\left(\frac{\partial^2\check  A}
{\partial p\,\partial a_n}\,\frac{\partial^2 \check B}
{\partial q \partial a^{*}_n}
-\frac{\partial^2 \check A}{\partial q\,\partial a_n}
\frac{\partial^2\check  B}
{\partial p\,
  \partial a^*_n}\,
\right)+\cdot\cdot\cdot .
\end{eqnarray}
This expansion leads to the $*$-bracket
\begin{eqnarray}\label{*-br}
\{{\check A},\,{\check B}\}_*= \frac{i}{\hbar}({\check A}*{\check B}-
{\check B}*{\check A})= \{{\check A},\,{\check B}\} +\hbar\, X_1({\check A},{\check B})
 +\hbar^2X_2({\check A},{\check B})+\cdot\cdot\cdot ,
\end{eqnarray}
which includes now the zero mode
contributions
\begin{eqnarray}\label{*-br-h}
X_1({\check A},{\check B})
=\frac{i}{2}\sum_{n,m>0}\,mn\left(  \frac{\partial^2{\check A}}
{\partial a_m\,\partial a_n}\,\,\frac{\partial^2{\check B}}
{\partial a^{*}_m\,\partial a^{*}_n}-
\frac{\partial^2{\check B}}
{\partial a_m\,\partial a_n}\,\,\frac{\partial^2{\check A}}
{\partial a^{*}_m\,\partial a^{*}_n}\right)+~~~~~~~~\nonumber \\
\frac{1}{2}\sum_{n>0}n\left(
\frac{\partial^2{\check A}}
{\partial p\,\partial a_n}\,\,\frac{\partial^2{\check B}}
{\partial q\,\partial a^{*}_n}-
\frac{\partial^2{\check B}}
{\partial p\,\partial a_n}\,\,\frac{\partial^2{\check A}}
{\partial q\,\partial a^{*}_n}+
\frac{\partial^2{\check A}}
{\partial p\,\partial a_n^*}\,\,\frac{\partial^2{\check B}}
{\partial q\,\partial a_n}-
\frac{\partial^2{\check B}}
{\partial p\,\partial a^*_n}\,\,\frac{\partial^2{\check A}}
{\partial q\,\partial a_n}\right).
\end{eqnarray}
$X_2({\check A},{\check B})$ contains third derivatives of
$\check A$ and $\check B$, {\it etc}.

\noindent
The symbols of the Hamiltonian (\ref{H_L}) and the generators of the
translations $\phi\,'(z)$ and dilatations $K=pq$ remain undeformed
\begin{eqnarray}\label{phi',H,K}
\check \phi\,' (z)=\phi\,' (z),~~~~~~
\check K= K,~~~~~~
\check H=H,
\end{eqnarray}
and their $*$-brackets with any $\check A$ then coincide, as expected,
with the Poisson brackets
 \begin{eqnarray}\label{*br-phi',H,K}
\{{\check A},\,\phi\,' (z)\}_*=
\{{\check A},\,\phi\,' (z)\},~~~~~
\{{\check A},\,K\}_*=
\{{\check A},\,K\},~~~~~~
\{{\check A},\,H\}_*=
\{{\check A},\,H\}.
\end{eqnarray}
This gives the realisation of the translation and
dilatation symmetries for the symbols.

\noindent
But we still  have to include the generators of the
conformal symmetry. We assume  that the commutation relations
corresponding to (\ref{PB-T-T,L}),
(\ref{PB-T-phi'}) are deformed at most by a central term.  Using
(\ref{PB-phi'-phi'}) and the Jacobi identity, the $*$-bracket for
(\ref{PB-T-phi'}) is determined up to a constant $C_0$
\begin{eqnarray}\label{*B-T-phi'}
&&\{\check{T} (z),\, \phi' (y)\}_*= \phi\,'' (y)\,\delta(z-y)-
\phi\,' (y)\,\delta\,'(z-y)+C_0\delta\,''(z-y).
\end{eqnarray}
Because its left-hand side coincides with
the Poisson bracket,  a first order
linear variational equation for $\check{T} (z)$
 arises which leads by integration to
\begin{equation}\label{checkT}
\check{T} (z)=\phi'\,^2\,(z)-C_0\,\phi''\,(z)+C_1(p,z),
\end{equation}
where the integration `constant' $C_1$ depends on $p$ and
$z$ only. Since (\ref{PB-T-T,L}) can be
deformed  by a central term only,  $C_1=0$, and  the
symbol of the energy-momentum tensor is
\begin{equation}\label{s-T-d}
   {\check T}(z) =\phi'\,^2\,(z)-C_0\,\phi''\,(z).
\end{equation}
It might be suggestive not to deform the generators of the conformal
symmetry \cite{Neveu, OW}, as it is the case for the group
$G(\rho(z),\epsilon)$, but this is not a necessary requirement for the
energy-momentum tensor \cite{Thorn, OW} and we take $\eta=\gamma
C_0$ as a deformation parameter for (\ref{T-phi})
\begin{equation}\label{s-T}
   {\check T}(z) =\phi'\,^2\,(z)-\frac{\eta}{\gamma}\,\phi''\,(z).
\end{equation}
Its relation to the undeformed expression is simply given by
$\gamma\mapsto \gamma\eta$, but for the time being $\eta$ is not
determined.

\noindent
Since  $\check T(z)$ has quadratic and linear terms in $\phi(z)$ only
its $*$-bracket with any $\check A$ does
not have terms of order higher than $\hbar$
\begin{eqnarray}\label{*-br-T}
\{\check T(z),\,{\check A}\}_*=
\{\check T(z),\,{\check A}\} +\hbar\, X_1(\check T(z),{\check A}),
\end{eqnarray}
where according to (\ref{*-br-h})
\begin{eqnarray}\label{*-br-T-h}
X_1(\check T(z),{\check A})
=\frac{i}{4\pi}\sum_{n,m>0}\,mn\left(  e^{-i(n+m)z}
\,\frac{\partial^2{\check A}}
{\partial a^{*}_m\,\partial a^{*}_n}-
e^{i(n+m)z}\,\frac{\partial^2{\check A}}
{\partial a_m\,\partial a_n}\,\,\right)+~~~~~~~~\nonumber \\
\frac{1}{(4\pi)^{3/2}}\sum_{n>0}n\left(e^{-inz}
\,\frac{\partial^2{\check A}}
{\partial q\,\partial a^{*}_n}+e^{inz}
\,\frac{\partial^2{\check A}}
{\partial q\,\partial a_n}\right).~~~~~~~~~~~~~~~~~~~~~~~~~~~~~~
\end{eqnarray}
Taking now ${\check A}=\check T(y)$ we obtain
\begin{equation}\label{B4}
X_1(\check T(z),{\check T(y)})=\frac{i\hbar}{8\pi^2}
\sum_{k\neq 0,\pm 1} e^{-ik(z-y)}\sum_{m=1}^{k-1}m\,(k-m)\,,
\end{equation}
 and using
\begin{equation}\label{B5}
\sum_{m=1}^{k-1}m\,(k-m)=\frac{k^3-k}{6}\,\,,
\end{equation}
the well-known Virasoro algebra results
\begin{eqnarray}\label{*B-T-T}
\{\check{ T} (z),\check{ T}(y)\}_*=
\check{ T}\,'(y)\,\delta (z-y) -2\check{T}(y)\delta\,'(z-y)
~~~~~~~~~~~~\nonumber \\
+\left(\frac{\eta^2}{2\gamma^2}+\frac{\hbar}{24\pi}\right)\delta\,'''(z-y)+
\frac{\hbar}{24\pi}\delta\,'(z-y).
\end{eqnarray}

\noindent
This completes the discussion of the elements we need for the application
of the Moyal formalism to the Liouville theory.

\subsection{The construction of symbols for chiral fields}

In general a symbol differs from its classical counterpart.  The
symbols of fields will be constructed by their transformation
properties under the symmetries of the theory. This principle also
operates for canonical quantisation .

\noindent
Let us first consider the chiral field $\check \psi (z)$. The
$*$-bracket relation which corresponds to (\ref{PB-phi'-psi}) is a
first order linear homogeneous variational equation for $\check \psi
(z)$, which leads to $\check \psi (z)=C(p,z)\psi (z)$, where
$C(p,z)$ is the integration `constant'.  Commutation with the
Hamiltonian $\{H,\check\psi(z)\}_*=\check\psi\,'(z)$ yields
$\partial_z C(p,z)=0$. Thus,
\begin{equation}\label{s-psi}
   {\check\psi}(z) =C(p) \, \psi(z)=C(p) \,
e^{-{\mathnormal\gamma}\phi(z)}.
\end{equation}
The conformal weight $\Delta(\check\psi)$ of $\check\psi (z)$
is defined by its commutation with $\check T (z)$ via (\ref{*-br-T}),
(\ref{*-br-T-h})
\begin{eqnarray}\label{*B-T-psi}
\{\check T (z),\check\psi(y)\}_*=\check\psi\,'(y)\,\delta (z-y) +
\frac{1}{2}
\left(\eta +
\frac{\hbar\gamma^2}{4\pi}\right)\check\psi(y)\delta\,'(z-y),
\end{eqnarray}
and we read off
\begin{equation}\label{delta-psi}
\Delta (\check\psi) =-\frac{1}{2}\left(\eta +
\alpha\right),
\end{equation}
where
\begin{equation}\label{alpha}
\alpha =\frac{\hbar\gamma^2}{4\pi}.
\end{equation}
An arbitrary exponential $e^{\mu\gamma\phi}$ is also primary with
\begin{equation}\label{delta-e}
\Delta (e^{\mu\gamma\phi}) =\frac{1}{2}(\eta\mu -\alpha\mu^2)\,.
\end{equation}
The construction of the symbol $\check\chi(z)$ requires more labour.
Since the field $\chi(z) $ is given by  integration of the bilocal
field (\ref{bilocal}) which linearly transforms under the translations
(\ref{bilocal1}), it is convenient first to construct $\check
B(y,x)$.  Using the commutation relation
\begin{equation}\label{bil-phi'}
\{\phi\,'(z), \check B(y,x)\}_*=-\frac{\gamma}{2}\delta(z-y)\,
\check B(y,x)+
\gamma \delta(z-x)\,\check B(y,x),
\end{equation}
 we find by the same
technique as given before that $\check B(y,x)=f(y,x;p)B(y,x)$
with an arbitrary function $f(y,x;p)$. The commutation of $\check B$
with the Hamiltonian
$\{H,\check B(y,x)\}_*=\partial_y\check B(y,x)+\partial_x\check B(y,x)$
leads to $f(y,x;p)=f(y-x;p)$.
To find the function
$f(y-x,p)$ we
commute $\check T(z)$ with
 \begin{equation}\label{s-chi}
{\check\chi}(y)= e^{-\gamma\phi (y)}\,\int_0^{2\pi}\,
dy\,f(y-x,p)\,\theta_{\gamma p}\, (y-x)\,e^{2\gamma\phi(x)}\,\,.
\end{equation}
For the detailed calculations we refer to Appendix C.
Eqs. (\ref{*-br-T}) and (\ref{*-br-T-h}) provide for ${\check
  A}=\check\chi (y)$ the result (\ref{*br-T-chi1}) with
anomalies. We require cancellation of these anomalies  in order to have
a primary  $\check \chi(y)$
\begin{equation}\label{*br-T-chi}
\{\check T (z),\check\chi(y)\}_*=\check\chi\,'(y)\,\delta (z-y) +
\frac{1}{2}
\left(\eta +\alpha\right)\check\chi(y)\delta\,'(z-y).
\end{equation}
This procedure determines uniquely the deformation parameter of the
energy-momentum tensor (\ref{s-T})
\begin{equation}\label{eta}
\eta =1 +{2\alpha}.
\end{equation}
For undeformed $\check T (z)$, given by $\gamma \rightarrow
\gamma\eta$, the $\eta$ is defined by the known quadratic
equation.

\noindent
Furthermore, the function $f(y-x,p)$ has to fulfil the first order
differential equation
\begin{equation}\label{f_alpha}
\partial_yf(y,p)=\alpha\cot (y/2)\,f(y,p),
\end{equation}
which has the solution
\begin{equation}\label{f-gamma}
f_\alpha (y) =\left(4\sin^2\frac{y}{2}\right)^{\alpha}.
\end{equation}
The field $\check\chi(y)$ then becomes
\begin{equation}\label{s-chi1}
{\check\chi}(y)=S(p)\,e^{-\gamma\phi (y)}\,\int_0^{2\pi}\,
dy\,\,e^{\frac{\gamma p}{2}\epsilon(y-x)}\,
f_\alpha (y-x)\,e^{2\gamma\phi(x)}\,\,,
\end{equation}
with the still arbitrary function $S(p)$.
The conformal weights of $\check\chi (z)$ and $\check\psi (z)$
have  quantum corrections,  but their weights remain equal
\begin{equation}\label{delta-psi-chi}
\Delta (\check \psi) = \Delta (\check \chi) =
-\frac{1}{2}\left(1+{3\alpha}\right).
\end{equation}
This provides primariness of
the Liouville field too.

\noindent
The symbol of the integral $A(z)$ (\ref{A(z)})  can be constructed
similarly and we obtain
\begin{equation}\label{s-A}
   {\check A}(z) =A(z),~~~~~~~~~\Delta(\check A)=0,
\end{equation}
which is consistent with
$\Delta(e^{2\gamma\phi})=1$.

\noindent
In Appendix D we will argue that the functions $C(p)$
of (\ref{s-psi}) and $S(p)$ of (\ref{s-chi1}) can be calculated
from dilatation properties of normalised vacuum Berezin symbols.

\noindent
Then the results found so far by the Moyal
quantisation are in agreement with those of the canonical quantisation
of the Liouville theory \cite{Thorn, OW}.

\subsection{The $*$-product exchange algebra of chiral fields}

In this section we calculate the $*$-products of the chiral fields
$\check\psi(z)$ and $\check\chi(z)$ to get the quantum exchange
algebra.  The symbols of $\psi(z)$ and $\chi(z)$ have the useful
property that the non-zero modes appear only linearly in the exponent.
The $*$-products of these symbols can therefore be calculated  directly
by Gaussian functional integration of (\ref{*-pr}). These
integrations give the results
\begin{eqnarray}\label{*pr-psi-psi}
&&\check\psi (z)*\check\psi(y)=
\,\psi (z)\,\psi(y)\,C_+(p)C_-(p)\,e^{-i\pi\alpha\epsilon^+(z-y)}\,,\\
\label{*pr-chi-chi}
&&\check\chi (z)*\check\chi(y)=
\,\psi (z)\,\psi(y)\,
I_0(z,y)\,e^{-i\pi\alpha\epsilon^+(z-y)}\,,\\
\label{*pr-psi-chi}
&&\check\psi (z)*\check\chi(y)=
\,\psi (z)\,\psi(y)\,I_1(z,y)\,e^{i\pi\alpha\epsilon^-(z-y)}\,,\\
\label{*pr-chi-psi}
&&\check\chi (z)*\check\psi(y)=
\,\psi (z)\,\psi(y)\,
I_2(z,y)\,e^{i\pi\alpha\epsilon^-(z-y)}\,,
\end{eqnarray}
where $C_\pm(p)=C(p\pm\hbar\gamma/2)$, and
$I_0$, $I_1$, and $I_2$ are the integrals
\begin{eqnarray}\label{I-chi-chi}
&&I_0(z,y)=S_+(p)\,S_-(p)
\int_0^{2\pi}\int_0^{2\pi}dx\,dv\,f_\alpha^{-2}(x-v)\,
f_\alpha(z-x)\,f_\alpha(y-v)\nonumber\\
&&~~~~~~~~~~
\times f_\alpha(z-v)\,f_\alpha(y-x)\,
\,e^{\frac{\gamma p}{2}\epsilon(z-x)}e^{\frac{\gamma p}{2}
\epsilon(y-v)}
\,e^{2\gamma\phi(x)+2\gamma\phi(v)},\\
\label{I-psi-chi-I}
&&I_1(z,y) =R_-(p)
\int_0^{2\pi}dx\,\,f_\alpha(z-x)\,f_\alpha(y-x)\,
e^{\frac{\gamma p}{2}\epsilon (y-x)}
\,e^{2\gamma\phi(x)}\,e^{-i\pi\alpha\kappa(z,y,x)}\,,\\
\label{I-chi-psi}
&&I_2(z,y)=R_+(p)\int_0^{2\pi}dx\,\,f_\alpha(z-x)\,f_\alpha(y-x)\,
\,e^{\frac{\gamma p}{2}\epsilon (z-x)}\,
e^{2\gamma\phi(x)}\,e^{-i\pi\alpha\kappa(z,y,x)}\,,
\end{eqnarray}
with
\begin{eqnarray}\label{kappa}
&&S_\pm(p)=S(p\pm\hbar\gamma/2),~~~~~~ R_\pm(p)=C_\pm(p) S_\pm(p),\nonumber \\
&&\kappa(z,y,x) =\epsilon(z-y)+\epsilon(y-x)+\epsilon(x-z),
\end{eqnarray}
and
\begin{equation}\label{eps_pm}
\epsilon^{\pm}(z)=
\frac{z}{2\pi}\pm\frac{i}{\pi}\sum_{n>0}\frac{e^{\pm inz}}{n}.
\end{equation}
The $\epsilon^{\pm}(z)$ functions are the positive and negative
frequency parts of the stair-step function
$\epsilon(z)=\epsilon^+(z)+\epsilon^-(z)$. They are related by
$\epsilon^+(-z)=-\epsilon^-(z)$ and contain the short distance
singularities
\begin{equation}\label{eps_pm-eps}
\epsilon^{\pm}(z)=\frac{1}{2}
\epsilon(z)\mp \frac{i}{2\pi}\log\left(4\sin^2\frac{z}{2}\right).
\end{equation}
We should mention here that the integration over the singularity
$f_\alpha^{-2}$ of (\ref{I-chi-chi}) is defined only if $\alpha <
1/4$, which restricts the coupling constant
$\gamma^2<\pi/\hbar$.

\noindent
The bilocal objects $I_0(z,y)$, $I_1(z,y)$ and $I_2(z,y)$ have the following
symmetry properties under the exchange of the $z$, $y$ coordinates
\begin{eqnarray}\label{exchange!_0}
&&I_0(y,z)=I_0(z,y),\\
\label{exchange!_1}
&&I_1(y,z)= a(p)I_2(z,y)+b(p)I_1(z,y)e^{\frac{\gamma p}{2}\epsilon(z-y)},\\
\label{exchange!_2}
&&I_2(y,z)= c(p)I_2(z,y)+d(p)I_1(z,y)e^{-\frac{\gamma p}{2}\epsilon(z-y)}.
\end{eqnarray}
Equation (\ref{exchange!_0}) is obvious.
To get (\ref{exchange!_1}) we take into account that the function
$\kappa (z,y,x)$ has in the integration region $x\in[0,2\pi]$
the two values $\kappa =\pm 1$ only.  Then the
equality of the integrands on the left and right hand sides define
the coefficients $a(p)$ and $b(p)$
\begin{equation}\label{a(p)}
a(p)=\frac{R_-(p)\,\sinh\left(\frac{1}{2}\gamma p
-2i\pi\alpha\right)}
{R_+ (p)\,\sinh\frac{1}{2}\gamma p~~~~~~~~~~~~},
~~~~~~b(p)=\frac{i\sin 2\pi\alpha}{\sinh\frac{1}{2}\gamma p},
\end{equation}
 The derivation of
(\ref{exchange!_2}) is similar and it yields
\begin{equation}\label{c(p)}
c(p)=\frac{R_+(p)\,\sinh\left(\frac{1}{2}\gamma p
+2i\pi\alpha\right)}
{R_- (p)\,\sinh\frac{1}{2}\gamma p~~~~~~~~~~~~},
~~~~~~d(p)=-\frac{i\sin 2\pi\alpha}{\sinh\frac{1}{2}\gamma p}.
\end{equation}
The function $R(p)=C(P)\,S(p)$ is given by (\ref{C.S})
\begin{equation}\label{R(p)}
R(p)=R_0\,\left(\sinh^2\frac{\gamma p}{2}+
\sin^2\pi\alpha\right)^{-\frac{1}{2}},
\end{equation}
which determines  $a(p)$ and $c(p)$ uniquely
\begin{equation}\label{a,c(p)}
a(p)=c(p)=\left(1+\frac{\sin^2(2\pi\alpha)}{\sinh^2\frac{\gamma p}{2}}.
\right)^{\frac{1}{2}}
\end{equation}
The quantum exchange algebra \cite{Neveu, OW} follows then immediately
from (\ref{*pr-psi-psi})-(\ref{*pr-chi-chi}) and
(\ref{exchange!_1})-(\ref{exchange!_0})
\begin{eqnarray}\label{*pr1-psi-psi}
e^{i\pi\alpha\epsilon(z-y)}\,\check\psi(z)*\check\psi(y)=
\check\psi (y)*\check\psi(z),
~~~~~~~~~~~~~~~~~~~~~~~~~~~~~~~~~~~~~~~~~~~~~~~~~~~~~~\\
\label{*pr1-chi-chi}
e^{i\pi\alpha\epsilon(z-y)}\,\check\chi(z)*\check\chi(y)=
\check\chi (y)*\check\chi(z),
~~~~~~~~~~~~~~~~~~~~~~~~~~~~~~~~~~~~~~~~~~~~~~~~~~~~~~\\
\label{*pr1-psi-chi}
e^{-i\pi\alpha\epsilon(z-y)}\,\check\psi(z)*\check\chi(y)=
\check\chi(y)*\check\psi(z)\,a(p)
-2i\sin (2\pi\alpha)\,
\check\psi(y)*\check\chi(z)
\,\theta_{-\gamma p}(z-y),\\
\label{*pr1-chi-psi}
e^{-i\pi\alpha\epsilon(z-y)}\,\check\chi(z)*\check\psi(y)=
\check\psi(y)*\check\chi(z)\, c(p)
-2i\sin (2\pi\alpha)\,
\check\chi(y)*\check\psi(z)\,\theta_{\gamma p}(z-y).~~~
\end{eqnarray}
Because $\check\psi(y)*\check\chi(z)$ does not depend on $q$ its
product with $a(p)$, $c(p)$ and $\theta_{\pm\gamma p}$ is here an
ordinary one.  Expanding (\ref{*pr1-psi-psi})-(\ref{*pr1-chi-psi}) in
powers of $\hbar$ the zero and first order terms in $\hbar$ reproduce
 the classical exchange algebra
(\ref{PB-psi-psi})-(\ref{PB-psi-chi}).

\subsection{The non-equal time $*$-brackets of Liouville  fields}

The symbol calculus for the anti-chiral part is similar.
Combining the chiral and anti-chiral fields we construct
the symbol of the Liouville exponential (\ref{Liouville})
\begin{equation}
\check u(z,\bar z)=\check E(z,\bar z)+\check K(z,\bar z),
\end{equation}
where
\begin{equation}\label{E,K}
\check E(z,\bar z)=\check \psi(z)\check {\bar\psi}(\bar z),~~~~~~~~~~~~
\check K(z,\bar z)=\check\chi(z)\check{\bar\chi}(\bar z).
\end{equation}
For the calculation of the $*$-product we treat the chiral and
anti-chiral fields independently  by using again the
Fubini-Veneziano trick and put $ p_0=\bar p_0=p$ and $q_0+\bar q_0=q$
afterwards
\begin{equation}\label{p=p_0}
A(p, p)e^{\alpha q}*B(p,p)e^{\beta q}=
A(p_0,\bar p_0)e^{\alpha(q_0+\bar q_o)}*
B(p_0,\bar p_0)e^{\beta(q_0+\bar q_o)}|_{p_0=\bar p_0=p, q_0+\bar q_0=q}.
\end{equation}
This relation obviously follows from (\ref{*pr0}).
The eqs. (\ref{*pr-psi-psi})-(\ref{*pr-chi-chi}) provide then
\begin{eqnarray}\label{E*E}
&&\check E(z,\bar z)*\check E(y,\bar y)=E(z,\bar z)E(y,\bar y)\,
C_+^2(p)C_-^2(p)e^{-i\pi\alpha[\epsilon^+(z-y)+\epsilon^+(\bar z-\bar y)]}\,\,, \\
\label{K*K}
&&K(z,\bar z)*K(y,\bar y)=E(z,\bar z)E(y,\bar y)
I_0(z,y)\bar I_0(\bar z,\bar y)\,
e^{-i\pi\alpha[\epsilon^+(z-y)+\epsilon^+(\bar z-\bar y)]}\,\,,\\
\label{E*K}
&&E(z,\bar z)*K(y,\bar y)=E(z,\bar z)E(y,\bar y)
I_1(z,y)\bar I_1(\bar z,\bar y)\,
e^{i\pi\alpha[\epsilon^-(z-y)+\epsilon^-(\bar z-\bar y)]}\,\,,\\
\label{K*E}
&&K(z,\bar z)*E(y,\bar y)=E(z,\bar z)E(y,\bar y)
I_2(z,y)\bar I_2(\bar z,\bar y)\,
e^{i\pi\alpha[\epsilon^-(z-y)+\epsilon^-(\bar z-\bar y)]}\,\,.
\end{eqnarray}
Our aim  is to find the quantum realisation of the non-equal
time Poisson brackets discussed in subsection 3.4. We consider
here for simplicity the case (\ref{PB-Liouville3}) only.
One expects the quantum relation
\begin{eqnarray}\label{u*u=}
 \check u(z,\bar z)*\check u(y,\bar y)=
C[\check u(z,\bar y)*\check u(y,\bar z)+
\check u(y,\bar z)*\check u(z,\bar y)]+
D \check u(y,\bar y)*\check u(z,\bar z),
\end{eqnarray}
where $C$ and $D$ are functions of space-time coordinates only. In order to
determine $C$ and $D$ we use the fact that the product $E(z,\bar
z)E(y,\bar y)$ is invariant under the exchanges of $z$ and $y$ or
$\bar z$ and $\bar y$. The comparison of the coefficients of $E*E$ on
the left and right hand sides of (\ref{u*u=}) gives the condition
\begin{eqnarray}\label{E*E,K*K}
\left(e^{i\pi\alpha\epsilon}+
e^{i\pi\alpha\epsilon}\right)C+
e^{i\pi\alpha\Theta}\,D=1.
\end{eqnarray}
Here we use the shorthand notation $\epsilon=\epsilon(z-y)$,
$\bar\epsilon=\epsilon(\bar z-\bar y)$, and $\Theta$ is defined by
(\ref{Theta}).  Due to the symmetry (\ref{exchange!_0}) the same
condition also holds for the coefficients of $K*K$. For the term
$E*K+K*E$ we use the exchange relations (\ref{exchange!_1}) and
(\ref{exchange!_2}) and rewrite the left and right hand sides of
(\ref{u*u=}) as a linear combination of the
four independent terms $I_1(z,y)\bar
I_1(\bar z,\bar y)$, $I_2(z,y)\bar I_2(\bar z,\bar y)$, $I_1(z,y)\bar
I_2(\bar z,\bar y)$ and $I_2(z,y)\bar I_1(\bar z,\bar y)$, which are
multiplied by $p$-dependent coefficients. In this manner we find four
equations but due to (\ref{a,c(p)})
only two of them are independent
\begin{eqnarray}\label{E*K+K*E}
&&\left(e^{-i\pi\alpha\epsilon}+
e^{-i\pi\alpha\bar\epsilon}\right)C+2i\sin(2\pi\alpha)\,\Theta
\,e^{-2i\pi\alpha\Theta}D=0,\\
\label{E*K+K*E1}
&&i\sin(2\pi\alpha)\left[\epsilon e^{-i\pi\alpha\epsilon}
+\bar\epsilon\,e^{-i\pi\alpha\bar\epsilon}\right]\,C
+[1-\sin^2(2\pi\alpha)(1+\epsilon\bar\epsilon)]
e^{-2i\pi\alpha\Theta}\,D=1.
\end{eqnarray}
Equations (\ref{E*E,K*K}) and (\ref{E*K+K*E}) determine
\begin{eqnarray}\label{C,D}
&&C=\frac{1}{e^{i\pi\alpha\epsilon}+e^{i\pi\alpha\bar\epsilon}}\,\,
\frac{2i\sin(2\pi\alpha)\Theta}{2i\sin(2\pi\alpha)\Theta -
e^{2i\pi\alpha\Theta}}\,\,,
\nonumber \\
&&D=-\frac{1}{2i\sin(2\pi\alpha)\Theta - e^{2i\pi\alpha\Theta}}\,\,,
\end{eqnarray}
which satisfy (\ref{E*K+K*E1}) as well.
The quantum non-equal time relation  becomes
\begin{eqnarray}\label{u*u}
\left[e^{2i\pi\alpha\Theta}-
\left(e^{2i\pi\alpha}-e^{-2i\pi\alpha}\right)\Theta \right]
\check u(z,\bar z)*\check u(y,\bar y)=
\check u(y,\bar y)*\check u(z,\bar z)~~~~~~~~~~~~~~~~~~~~~~~~\\ \nonumber
-\Theta\,\frac{e^{2i\pi\alpha}-e^{-2i\pi\alpha}}{e^{i\pi\alpha\epsilon}+
e^{i\pi\alpha\bar\epsilon}}
[\check u(z,\bar y)*\check u(y,\bar z)+
\check u(y,\bar z)*\check u(z,\bar y)].
\end{eqnarray}
It provides for $\Theta =0$  obviously the causality condition
$\{\check u(z,\bar z),\,\check u(y,\bar y)\}_*=0$.
For $\Theta =\pm 1$ (\ref{u*u}) relates quadratic combinations
of the field $u$  at  four different space-time points,
which can also be written as a $*$-bracket, and combining it
with the case $\Theta=0$ we get the final result
\begin{eqnarray}\label{u*u!}
\{\check u(z,\bar z),\,\check u(y,\bar y)\}_*=
\frac{1}{\hbar}\,\sin(\hbar\gamma^2/4)\,
[\epsilon(z-y)+\epsilon(\bar z-\bar y)]\times
~~~~~~~~~~~~~~~~~~~~~
\\ \nonumber
\,\left[\check u(z, \bar y) * \check u(y, \bar z) +\check u(y, \bar z)*
\check u(z, \bar y)-
\frac{\check u(z, \bar z)* \check u(y, \bar y)+\check u(y, \bar y)*
\check u(z, \bar z)}{2\cos(\hbar\gamma^2/4)}
\right].
\end{eqnarray}
Much in the same manner as for the derivation of (\ref{canonical-PB})
we can easily construct from it a symbol of the local field
$e^{-2\gamma\phi(\tau,\sigma)}$.

\noindent
Its expansion in powers of $\hbar$ reproduces the classical
Poisson bracket (\ref{PB-Liouville3}).

\section{Summary}
In this paper the Liouville theory was revisited, classically and
quantum mechanically. Its Poisson and symmetry structures are defined
from the $SL(2,\rr)$ WZNW theory by gauge invariant Hamiltonian
reduction, and for quantisation a Moyal formalism is applied.

\noindent
The classical form of the exchange algebra arises as the basic Poisson
algebra, from which causal non-equal time Poisson brackets for a
Liouville exponential are derived.  We observed a transitive symmetry
group acting on the phase space, and we have shown that Hamiltonian
reduction is a suitable method for the integration of gauged WZNW
theories.

\noindent
Following the ideas of geometric quantisation, coherent states have
been defined via this transitive symmetry group and a symbol calculus
developed. The symbols of fields are constructed through the
symmetries of the Liouville theory.
We reproduce results of
the canonical quantisation, which includes the deformed exchange
algebra. In addition, the deformed causal non-equal time commutators
of a Liouville exponential is calculated.
From this rich structure other symbols and $*$-products can be derived.

\noindent
We presume that vacuum Berezin symbols provide a natural
definition for Liouville correlation functions.

\vspace{0.8 cm}
\noindent
{\bf {\Large Acknowledgements}}

\vspace{0.3cm}

\noindent
We would like to thank Chris Ford for many helpful discussions.
We also thank Martin Reuter for interesting discussions about
the Moyal formalism.
G.J. is grateful to DESY Zeuthen for hospitality.  His
research was supported by grants from the DFG, GSRT INTAS and RFBR.

\vspace{0.5cm}
\vspace{0.5cm}

\setcounter{equation}{0}
\def\theequation{A.\arabic{equation}}

{\bf {\Large Appendix A}}

\vspace{0.5cm}

\noindent
For Poisson bracket calculations the technique of symplectic geometry
is applied \cite{Woodhouse}.  If a symplectic form is given by
$\omega=\omega_{mn}(\xi)\, \delta \xi^m\wedge\delta\xi^n$, then for
an observable $F(\xi)$ one has
\begin{equation}\label{PB-A}
-\delta F =2\omega_{mn}\,(\xi)\,\,\{\,F,\xi^m\,\}\,\delta\xi^n ,
\end{equation}
which for non-degenerate $\omega_{mn}\,(\xi)$ defines the Poisson brackets
$\{\,F,\xi^m\,\}$.

\noindent
We make use of (\ref{PB-A}) for the symplectic form (\ref{omega-WZ})
(for a more general treatment see \cite{BFP}). First we parametrise the chiral
field $g(z)$ by
\begin{equation}\label{g-f}
g(z)=f(z)\,M(z),~~~~~~
M(z)=\exp{\left(\frac{\lambda z}{2\pi}\,\,t_2\right)},
\end{equation}
where $f(z)$ is a $SL(2,\rr)$ valued periodic field.
This parametrisation assumes
 the monodromy matrix  $M=\exp{\lambda t_2}$. Then, from
(\ref{omega-WZ}) we get
\begin{eqnarray}\label{omega-f}
\omega =-\frac{1}{\gamma^2}\,
\int_0^{2\pi}\,\, \langle(f^{-1}(\sigma)\,\delta f(\sigma))'\wedge\,
f^{-1}(\sigma)\,\delta f(\sigma)\rangle\,d\sigma \nonumber \\
-\frac{\lambda}{2\pi\gamma^2}\,\int_0^{2\pi}d\sigma\,\,
\langle [f^{-1}(\sigma)\delta f(\sigma), t]\, \wedge\,
f^{-1}(\sigma)\delta f(\sigma)\rangle\nonumber \\
-\frac{1}{\pi\gamma^2}\,\,
\delta\lambda \wedge\langle t\int_0^{2\pi}d\sigma\,\,
f^{-1}(\sigma)\delta f(\sigma)\rangle.
\end{eqnarray}
 For $F=\lambda$ we read off from  (\ref{omega-f}) the equations
$$ A'(\sigma) +\frac{\lambda}{2\pi} [A(\sigma),t_2]=0~~~~
\mbox{and}~~~ \frac{1}{\pi\gamma^2}\int_0^{2\pi}d\sigma\,
\langle t_2~ A (\sigma)\rangle =-1,$$
where $A(\sigma) =
f^{-1}(\sigma)\,\{\lambda, f(\sigma)\}$. Due to the
 periodicity of $f(\sigma)$ we get the unique solution
\begin{equation}\label{PB-f-lambda}
\{\lambda ,f(\sigma)\} =\frac{\gamma^2}{2}\,f(\sigma)\,t_2.
\end{equation}
In order to write down the Poisson brackets of group elements
it is convenient to introduce the notation \cite{Goddard}
\begin{equation}\label{PB-fxf}
\{\,f_{\alpha_1\beta_1}(\sigma_1),\,f_{\alpha\beta}(\sigma)\,\}=
\{\,f(\sigma_1)\otimes f(\sigma)\,\}_{\alpha_1\alpha,\,\beta_1\beta}\,.
\end{equation}
For $F=f_{\alpha_1\beta_1}(\sigma_1)$,
the Poisson brackets of $f(\sigma)$'s can then be written as
\begin{equation}\label{PB-f-f}
\{\,f(\sigma_1),\otimes f(\sigma)\,\}=
A^{(n)}(\sigma_1,\sigma)\otimes
(f(\sigma)\,\,t_n),
\end{equation}
where the $t_n$ are given by (\ref{T}), and the
$A^{(n)}$ satisfy the equations
\begin{eqnarray}\label{eq-A^n}
\partial_\sigma A^{(n)}(\sigma_1,\sigma)+
\frac{\lambda}{\pi}\,\epsilon^n\,_{m2}\,
A^{(m)}(\sigma_1,\sigma) -
\frac{\gamma^2}{4\pi}\,\delta^n_2\, (f(\sigma_1)\,t_2)=
\nonumber \\
\frac{\gamma^2}{2}\, (f(\sigma_1)\,t^n)\,
\delta (\sigma_1-\sigma),
\end{eqnarray}
and
\begin{equation}\label{intA2=0}
\int_0^{2\pi}d\sigma\,
 A^{(2)}(\sigma_1,\sigma) = 0.
\end{equation}
Eq. (\ref{eq-A^n}) splits into independent equations
\begin{eqnarray}\label{eq-A2}
\partial_\sigma A^{(2)}(\sigma_1,\sigma)=
-\frac{\gamma^2}{2}\, (f(\sigma_1)\,t_2)
\left(\delta(\sigma_1-\sigma)-\frac{1}{2\pi}\right),\\
\label{eq-Apm}
\partial_\sigma A^{(\pm)}(\sigma_1,\sigma) \pm
\frac{\lambda}{\pi}A^{(\pm)}(\sigma_1,\sigma)=
-\frac{\gamma^2}{2}\, (f(\sigma_1)\,t_\mp)\,
\delta (\sigma_1-\sigma),
\end{eqnarray}
where $A^{(\pm)}=A^1\pm A^0$.
Eq. (\ref{intA2=0})-(\ref{eq-Apm}) have an unique solution
for $\lambda \neq 0$, which inverts the symplectic
form. Indeed, from (\ref{eq-A2}) and (\ref{intA2=0}) we find
\begin{equation}\label{A2}
A^{(2)}(\sigma_1,\sigma)=
\frac{\gamma^2}{4}\,\, (f(\sigma_1)\,t_2)\,\,
h(\sigma_1-\sigma),
\end{equation}
and integration of (\ref{eq-Apm}) gives
\begin{equation}\label{Apm}
A^{(\pm)}(\sigma_1,\sigma)=
\frac{\gamma^2}{2}\, (f(\sigma_1)\,t_\mp)\,\,
h_{\mp\lambda} (\sigma_1-\sigma),
\end{equation}
where
\begin{equation}\label{h_lambda}
h_\lambda (z)=\frac{e^{\lambda h(z)}}{2\,\sinh\lambda}.
\end{equation}
Note that for $\lambda\neq 0$
the operator $\partial_\sigma +\lambda/\pi$ has an inverse
on the class of periodic functions, and
$h_\lambda (\sigma -\sigma_1)$ is its kernel.

\noindent
Now using  (\ref{PB-f-f}) and (\ref{A2}), (\ref{Apm}) we obtain
\begin{eqnarray}\label{PB-f}
\{\,f(\sigma_1)\otimes f(\sigma)\,\}=
\frac{\gamma^2}{4}\,[\,(f(\sigma_1)\,t_2)\otimes
(f(\sigma)\,t_2)\,\,h(\sigma_1-\sigma) \nonumber\\
+(f(\sigma_1)\,t_-)\otimes
(f(\sigma)\,t_+)~
h_{-\lambda} (\sigma_1-\sigma)\nonumber\\
+\, (f(\sigma_1)\,t_+)\otimes
(f(\sigma)\,t_-)~
h_{+\lambda} (\sigma_1-\sigma)].
\end{eqnarray}
With this result we are able to calculate the Poisson brackets
of the chiral WZNW fields
\begin{eqnarray}\label{PB-gxg}
\{\,g(z_1)\otimes g(z)\,\}=
\{\,f(z_1)\otimes f(z)\,\}\cdot(\,M(z_1)\otimes M(z)\,)\nonumber\\
+(\,I\otimes f(z)\,)\cdot
\{\,f(z_1)\otimes M(z)\,\}\cdot(\,M(z_1)\otimes I\,)\nonumber\\
+(\,f(z_1)\otimes I\,)\cdot
\{\,M(z_1)\otimes f(z)\,\}\cdot(\,I\otimes M(z)\,)
\end{eqnarray}
explicitly. From (\ref{PB-f-lambda}) follows
\begin{eqnarray}\label{PB-f-M}
\{\,f(z_1)\otimes M(z)\,\} =-\frac{\gamma^2z}{4\pi}
\,(\,f(z_1)\,t_2\,)\otimes (\,M(z)\,t_2\,),\nonumber \\
\{\,M(z_1)\otimes f(z)\,\} =\frac{\gamma^2z_1}{4\pi}
\,(\,M(z_1)\,t_2\,)\otimes (\,f(z)\,t_2\,),
\end{eqnarray}
and it is easy to check that
\begin{equation}\label{M-t-M}
M^{-1}(z)\,t_\pm\,M(z)= \,t_\pm\,\,
\exp{\left(\pm\frac{\lambda z}{\pi}\right)}.
\end{equation}
Putting (\ref{PB-f})-(\ref{M-t-M}) into (\ref{PB-gxg})
we get our final result (\ref{PB-g}).

\vspace{0.3cm}
\noindent
Another remark is in order: because of (\ref{PB-f-lambda})
the Kac-Moody currents (\ref{KM-current}) have
zero Poisson brackets with $\lambda$. This simplifies the calculation of
$\{J_n(z_1),\,g(z)\,\}$ from (\ref{omega-WZ})
(with $M=\exp (\lambda t_2))$.  Eq. (\ref{PB-A}) gives for $F=J_n(z_1)$
\begin{equation}\label{delta-J}
-\delta J_n(z_1)=\frac{2}{\gamma^2}\,
\int_\tau^{\tau +2\pi}\,\, \langle(g^{-1}(z)\,\delta g(z))'\,
g^{-1}(z)\,\{\,J_n(z_1), g(z)\,\}\rangle\,dz,
\end{equation}
since the term proportional to $\delta\lambda$ is cancelled by
partial integration. The variation of (\ref{KM-current})
$$\delta J_n(z) = \frac{1}{\gamma^2}\, \langle \,t_n\,
g(z)\,((g^{-1}(z)\,\delta g(z))'\,g^{-1}(z)\,\rangle , $$
and (\ref{delta-J}) then provide (\ref{PB-J-g}).

\vspace{1cm}

\setcounter{equation}{0}
\def\theequation{B.\arabic{equation}}

{\bf {\Large Appendix B}}

\vspace{0.5cm}

\noindent
The coherent states on the half-plane $p>0$ are transformed into each other
by an irreducible representation of the group of translations and dilatations
\begin{equation}\label{T+D}
 q\mapsto q+b;~~~~~~p\mapsto  e^{-\varepsilon p},~~~~~~q\mapsto
e^{\varepsilon}q.
\end{equation}
In the complex coordinates
\begin{equation}\label{zeta}
\zeta=\frac{1}{p}+i\nu q,~~~~~~\zeta^*=\frac{1}{p}-i\nu q,
\end{equation}
these transformations become a  holomorphic map
$\zeta\mapsto  e^{-\varepsilon} \zeta-i\nu b$
of the half-plane onto itself.
(\ref{zeta}) is the analogue of (\ref{a,a+})
for the half-plane and $\nu$ is a squeezing parameter.
The coherent state $\Psi_{\zeta_1}\,(p)$ is defined as an eigenstate
of the operator
$\hat \zeta=1/p-\hbar\nu/2p-\hbar\nu\partial_p$
\begin{equation}\label{eigenv-z}
\left( \frac{1-\hbar\nu/2}{p}-\hbar\nu\partial_p\right)\Psi_{\zeta_1}\,(p)=
\zeta_1\Psi_{\zeta_1}\,(p),
\end{equation}
with the  eigenvalue $\zeta_1=1/p_1+i\nu q_1$.
We use here the standard measure $dp$ for the scalar product of
wave functions
\begin{equation}\label{scalar-pr}
\langle \Psi_{\zeta_1}|\Psi_{\zeta_2} \rangle=\int dp\,\Psi_{\zeta_1}^*(p)
\Psi_{\zeta_2}(p),
\end{equation}
and the term $\hbar\nu/2p$ of the operator
$\hat \zeta$ is necessary in order to
get the correct normalisation and completeness of the coherent states.
This term can also be justified with in
geometric quantisation \cite{Woodhouse} but
it will be cancelled if we pass to the dilatation invariant measure
$dp/p$.

\noindent
The solution of (\ref{eigenv-z}) with $\beta=1/\hbar\nu$ is
\begin{equation}\label{eigenstate-z}
\Psi_{\zeta_1}\,(p)=C_\beta(p_1,q_1)\,\frac{1}{\sqrt p}\,\,
p^{\beta}\,e^{-\beta\zeta_1p}
\end{equation}
and the integration constant $C_\beta (p_1,q_1)$
can be defined by the conditions
 \begin{equation}\label{T-D-Psi}
\hat p\Psi_{\zeta_1}\,(p)= i\hbar\partial_{q_1}\Psi_{\zeta_1}\,(p),
~~~~~\mbox{and}~~~~\hat K\Psi_{\zeta_1}\,(p)=i\hbar(q_1\partial_{q_1}
-p_1\partial_{p_1})\Psi_{\zeta_1}\,(p),
\end{equation}
which implies that $\hat p=p$ and
$\hat K=i\hbar(p\partial_p+1/2)$ are the translation and
dilatation operators respectively.
From (\ref{T-D-Psi}) follows
$C_\beta(p_1,q_1)=c_\beta\,p_1^{-\beta}$,
and the normalisation condition
$\langle \Psi_{\zeta_1}|\Psi_{\zeta_1}\rangle =1$ yields
$c_\beta^2 \,\Gamma(2\beta)\,=(2\beta)^{2\beta}$. Thus we get
the wave functions
\begin{equation}\label{coh-st}
\Psi_{\zeta_1}\,(p)=(2\beta)^{\beta}\Gamma^{-1/2}(2\beta)
\frac{1}{\sqrt p}\,\,
\left(\frac{p}{p_1}\right)^{\beta}\,e^{-\beta\zeta_1p}.
\end{equation}
which have the scalar product
\begin{equation}\label{scalar-pr1}
\langle \Psi_{\zeta_1}|\Psi_{\zeta_2} \rangle=(p_1p_2)^{-\beta}
\left(\frac{\zeta_1+\zeta^*_2}{2}\right)^{-2\beta},
\end{equation}
and satisfy the completeness relation
\begin{equation}\label{complete}
\left(1-\frac{\hbar\nu}{2}\right)
\int \frac{dp\,dq}{2\pi\hbar} \Psi_\zeta^* (p_1)\Psi_\zeta (p_2)=
\delta (p_1-p_2).
\end{equation}
This structure defines the
the Berezin symbol calculus for
\begin{equation}\label{Berezin}
\check A(p,q)=\langle \Psi_{\zeta}|\hat A|\Psi_{\zeta} \rangle.
\end{equation}

\noindent
At $\nu=0$ the coherent states become non-normalisable, like the standard
coherent states on the plane. In particular for $\nu\rightarrow 0$
we have
\begin{equation}\label{nu=0}
|\Psi_{\zeta_1}(p)|^2 \rightarrow \delta(p-p_1).
\end{equation}

\vspace{0.5cm}

\setcounter{equation}{0}
\def\theequation{C.\arabic{equation}}

{\bf {\Large Appendix C}}

\vspace{0.5cm}

\noindent
In this appendix we describe the cancellation of anomalies
needed to get a primary $\check\chi(y)$.
First we calculate the $*$-bracket of $\check T(z)$ with the
bilocal field
\begin{equation}
\check B(y,x)=F(y-x,p)\,e^{-\gamma\phi(y)}\,e^{2\gamma\phi(x)}
\end{equation}
taking into account
(\ref{*B-T-psi}), and (\ref{delta-e}) for $\mu =2$. Then eqs.
(\ref{*-br-T}), (\ref{*-br-T-h}) provide
\begin{eqnarray}\label{*b-T-B}
\{\check T (z),\,\check B(y,x)\}_*=
\left(e^{-\gamma\phi(y)}\right)'\,F(y-x,p) e^{2\gamma\phi(x)}\, \delta (z-y)
~~~~~~~~~~~~~\nonumber\\
+\frac{1}{2}\left(\eta +\alpha\right)
e^{-\gamma\phi(y)}\,F(y-x,p)\, e^{2\gamma\phi(x)}\,\delta\,' (z-y)+\nonumber\\
e^{-\gamma\phi(y)}\,F(y-x,p)\left( e^{2\gamma\phi(x)}\right)'\,\delta (z-x)\,
-\left(\eta -2\alpha\right)
e^{-\gamma\phi(y)}\,F(y-x,p)\, e^{2\gamma\phi(x)}\,\delta\,' (z-x)\nonumber\\
+\alpha\,e^{-\gamma\phi(y)}\,F(y-x,p)\, e^{2\gamma\phi(x)}
\cot\frac{1}{2}(y-z)\,[\delta (z-y)-\delta (z-x)],
\end{eqnarray}
where we have used
\begin{eqnarray}\label{*br-T-B1}
\frac{i}{2\pi}\sum_{k\geq 2} e^{-ik(z-y)}\sum_{m=1}^{k-1}e^{-im(y-x)}
+\frac{i}{4\pi}\sum_{k\geq 1}\left( e^{-ik(z-y)}+e^{-im(y-x)}\right)
+c.c.=\nonumber \\
\frac{1}{2}\cot\frac{1}{2}(y-z)\,[\delta (z-y)-\delta (z-x)].
\end{eqnarray}
The integration of (\ref{*br-T-B1}) over $x$ gives the result
\begin{eqnarray}\label{*br-T-chi1}
\{\check T (z),\check\chi(y)\}_*=\check\chi\,'(y)\,\delta (z-y) +\frac{1}{2}
\left(\eta +\alpha\right)\check\chi(y)\delta\,'(z-y) \nonumber\\
+\left(1-\eta+2\alpha\right)\,e^{-\gamma\phi(y)}
F(y-z,p)\,\left(e^{2\gamma\phi(z)}\right)'
\nonumber \\
+(\eta-2\alpha)\,e^{-\gamma\phi(y)}
F\,'(y-z,p)\,e^{2\gamma\phi(z)}
-\alpha\,e^{-\gamma\phi(y)}
\cot\frac{1}{2}(y-z)\,F (y-z,p)\,e^{2\gamma\phi(z)}\nonumber\\
-\left[e^{-\gamma\phi(y)}
\int_0^{2\pi}dx\,\left(F\,'(y-x,p)-
\alpha\cot\frac{1}{2}(y-x)\,F(y-x,p)\right)
\,e^{2\gamma\phi(x)}\right]\delta(z-y),
\end{eqnarray}
which has in addition to the standard terms undesirable anomalies.
The cancellation of the anomalies uniquely yields
(\ref{eta}) and (\ref{f_alpha}).

\vspace{0.5cm}
\setcounter{equation}{0}
\def\theequation{D.\arabic{equation}}

{\bf {\Large Appendix D}}

\vspace{0.5cm}

\noindent
Here we present some heuristic considerations and argue
that the equation
(\ref{R(p)}) can be derived by means of the dilatation properties
of the vacuum configuration (\ref{vacuum-symm}). For this
purpose let us define a new (s-)symbol
$A_s(p)e^{2\alpha q}$ with $A_s(p)>0$ and associate it with the operator
\begin{equation}\label{A_s}
\hat A =\sqrt{A_s(\hat p)}\,\,F_{2\alpha}(\hat p+i\hbar\alpha)
\,e^{2\alpha\hat q}
\,\,\sqrt{A_s(\hat p)}.
\end{equation}
The function $F_{2\alpha}(p)$ is included here in order to
get a  Hermitian operator
$\hat A$.
Such a symbol is related to the vacuum expectation value of $\hat A$,
and it can be obtained by a normalised limiting
procedure $\nu\rightarrow 0$ from the Berezin symbols (\ref{Berezin}).
Comparing the two different forms (\ref{hatA}) and (\ref{A_s})
of an operator $\hat A$,
the functions $A(p)$ and $A_s(p)$ are related by
\begin{equation}\label{A-A_s}
A(p) =\sqrt{A_s(p+i\hbar\alpha)\,A_s(p-i\hbar\alpha)}\,\,F_{2\alpha}(p).
\end{equation}
This relationship will now be used to specify the undetermined
functions $C(p)$ and $S(p)$ of (\ref{s-psi}) and (\ref{s-chi1}).
The dilatation properties of the vacuum configuration
(\ref{vacuum-symm}) give
for the s-symbol
\begin{eqnarray}\label{vacuum-psi_s,chi_s}
\check\psi_s(z)|_{q=0,\phi^\pm=0}=
c\,e^{-\frac{\gamma pz}{4\pi}},~~~~~
\check\chi_s(z)|_{q=0,\phi^\pm=0}=\frac{s}{p}\,
\,e^{\frac{\gamma pz}{4\pi}},
\end{eqnarray}
where $c$ and $s$ are constants, and
(\ref{A-A_s}) leads to
\begin{eqnarray}\label{vacuum-psi}
&&\check\psi(z)|_{q=0,\phi^\pm=0}=c\,F_{-\gamma}(p)
\,e^{-\frac{\gamma pz}{4\pi}},\\
\label{vacuum-chi}
&&\check\chi(z)|_{q=0,\phi^\pm=0}=s\,F_\gamma(p)\left[p^2+
(\hbar\gamma/2)^2\right]^{-\frac{1}{2}}\,
\,e^{\frac{\gamma pz}{4\pi}}.
\end{eqnarray}
But the vacuum configurations of  (\ref{s-psi}) and (\ref{s-chi1}) are
also given directly
\begin{eqnarray}\label{vacuum-psi1}
&&\check\psi(z)|_{q=0,\phi^\pm=0}=C(p)\,e^{-\frac{\gamma pz}{4\pi}} ,\\
\label{vacuum-chi1}
&&\check\chi(z)|_{q=0,\phi^\pm=0}=S(p)\,
\frac{A_\alpha\,e^{\frac{\gamma pz}{4\pi}}}{
\Gamma(1+\alpha+i\frac{\gamma p}{2\pi})
\Gamma(1+\alpha-i\frac{\gamma p}{2\pi})}.
\end{eqnarray}
To get (\ref{vacuum-chi1}) have used the integral
\begin{equation}\label{integral}
\int_0^{\pi}\,dx\, (\sin x)^{2\alpha}\,e^{\frac{\gamma px}{\pi}}=A_\alpha
\,\frac{e^{\frac{\gamma p}{2}}}{\Gamma
(1+\alpha+i\frac{\gamma p}{2\pi})\,\,
\Gamma(1+\alpha -i\frac{\gamma p}{2\pi})},
\end{equation}
where $A_\alpha$ is a $p$-independent constant.
The comparison of (\ref{vacuum-psi})-(\ref{vacuum-chi1}) relates
the unknown functions
\begin{equation}\label{C,S-F}
C(p)\,S(p)=\,\frac{F_{-\gamma}(p)\,F_{\gamma}(p)\,
\Gamma(1+\alpha-i\frac{\gamma p}{2\pi})\,
\Gamma(1+\alpha+i\frac{\gamma p}{2\pi})}
{[p^2+(\frac{\hbar\gamma}{2})^2]^{\frac{1}{2}}},
\end{equation}
where $R_0$ is a $p$-independent constant.
 The functions $F_\gamma$ can be calculated, in principle, by
a normalised limiting procedure of Berezin symbols
which we have not done yet.
Instead, we make here a suggestion which is guided by
locality consideration \cite{Thorn,OW}
\begin{equation}\label{F.F}
F_{-\gamma}(p)\,F_{\gamma}(p)=
\left(\frac{\Gamma(1-\alpha-i\frac{\gamma p}{2\pi})\,
\Gamma(1-\alpha+i\frac{\gamma p}{2\pi})}
{\Gamma(1+\alpha-i\frac{\gamma p}{2\pi})\,
\Gamma(1+\alpha+i\frac{\gamma p}{2\pi})}
\right)^\frac{1}{2}\,.
\end{equation}
The identity
\begin{equation}\label{Gamma.Gamma}
\Gamma(1-a-ib)\Gamma(1-a+ib)
\Gamma(1+a-ib)\Gamma(1+a+ib)=
\frac{(\pi a)^2+(\pi b)^2}{\sinh^2\pi a +\sin^2\pi b},
\end{equation}
provides finally
\begin{equation}\label{C.S}
C(p)\,S(p)=R_0\,\left(\sinh^2\frac{\gamma p}{2}+
\sin^2\pi\alpha\right)^{-\frac{1}{2}}.
\end{equation}
This result defines the
coefficients of the quantum exchange
algebra, and it  leads to the causal commutation relations
(\ref{u*u!}).

\end{document}